\documentclass[superscriptaddress,footinbib,prl,aps,twocolumn,floatfix]{revtex4-2}

\pdfoutput=1

\usepackage{slashed}
\usepackage{dcolumn}
\usepackage{bm}
\usepackage{xcolor}
\usepackage[utf8]{inputenc}
\usepackage[english]{babel}
\usepackage{amsmath,amssymb,amsfonts,latexsym,cancel}
\usepackage{graphicx}
\usepackage{braket}
\usepackage[colorlinks=true,linkcolor=red,urlcolor=blue,citecolor=blue]{hyperref}
\usepackage{orcidlink,booktabs}
\usepackage{siunitx,bm}
\usepackage{tikz}
\usepackage{comment}
\usetikzlibrary{arrows.meta, decorations.markings, decorations.pathreplacing, calligraphy, backgrounds, calc}

\newcommand{\mi}{\mathrm{i}}
\newcommand{\dbar}{\mathchar'26\mkern-12mu \mathrm{d}}
\newcommand{\deltabar}{\mathchar'26\mkern-9mu \delta}

\begin{document}

\noindent\hfill\texttt{CALT-TH/2026-009}\\[-10ex]

\title{Matter-wave Interferometers as Open-system Dark Matter Detectors}

\author{Leonardo Badurina}
\email{badurina@caltech.edu}
\affiliation{Walter Burke Institute for Theoretical Physics \& Leinweber Forum for Theoretical Physics, California Institute of Technology, Pasadena, CA 91125, USA}

\author{Kathryn M.\ Zurek}
\email{kzurek@caltech.edu}
\affiliation{Walter Burke Institute for Theoretical Physics \& Leinweber Forum for Theoretical Physics, California Institute of Technology, Pasadena, CA 91125, USA}

\date{\today}

\begin{abstract}
Matter-wave interferometers (MWIs) offer a uniquely quantum route to dark matter (DM) detection: DM can reveal itself through phase and decoherence between spatially separated wavepackets, even when energy deposition is negligible and no resolvable recoil displacement occurs. 
We formulate these effects in an open effective field theory for MWIs using the Schwinger--Keldysh formalism, which highlights a structural asymmetry between the two detection channels. 
For elastic spin-independent DM scattering, decoherence inherits novel Bose enhancement or Pauli blocking factors,  while the phase is at most linear in the DM occupation number. 
By retaining the DM's coherence time, this framework spans Markovian and non-Markovian dynamics across a wide range of DM masses, and systematically organizes corrections beyond the heavy-probe limit.

\end{abstract}

\maketitle

\textit{Introduction.}---Matter-wave interferometers (MWIs) could detect dark matter (DM) through channels that no classical detector can access. 
By placing atoms or mesoscopic objects in genuine quantum superposition, MWIs can be sensitive to a DM-induced phase shift and loss of contrast even in the absence of energy deposition or detectable recoil displacement. 
This sensitivity, together with their exceptional metrological precision, makes MWIs powerful probes of DM across an enormous mass range. 
Proposed atom gradiometers could detect linearly coupled and quadratically coupled scalar~\cite{Arvanitaki:2016fyj,Badurina:2021lwr,Badurina:2023wpk,Badurina:2019hst,Banerjee:2022sqg}, spin-2~\cite{Blas:2024kps} and axion-like particle (ALP)~\cite{Beadle:2023flm} ultralight DM in the $10^{-18}-10^{-12}$~eV range. 
Spin-superposition~\cite{Graham:2017ivz} and dual-species atom interferometers~\cite{Graham:2015ifn,MAGIS-100:2021etm,Zhou:2024qdw} could probe ALPs and vector DM, while MWIs employing mesoscopic objects could constrain sub-GeV DM~\cite{Riedel:2012ur,Riedel:2016acj,Du:2022ceh,Badurina:2024nge}. 
Space-based platforms may probe DM clumps gravitationally~\cite{Badurina:2025xwl}, and trapped-ion interferometers could detect magnetic fields sourced by ultralight ALP and dark photon DM~\cite{Badurina:2025idj}.

Despite this breadth, the theoretical description of MWIs in a DM background is fragmented. 
In the sub-GeV mass range one employs an $S$-matrix formalism to describe ultrasoft scattering off the detector, which assumes the Markovian regime and reproduces collisional decoherence~\cite{PhysRevA.42.38,Hornberger:2003umw}. 
For ultralight bosonic DM, ${m_{_\chi}\!\ll\!10}$~eV, DM is treated as a classical wave and all calculations proceed in the classical limit. 
Neither picture captures: (i) how the occupation number and statistics of DM imprint themselves on MWI observables, (ii) how these signatures interpolate across the wave-particle boundary, nor (iii) what happens when the DM coherence time exceeds the interferometric sequence and the dynamics become non-Markovian~\cite{Breuer:2007juk}.

In this \textit{Letter}, we address these points by treating a MWI as an \textit{open} quantum system coupled to a DM environment. 
Even though the joint MWI-plus-DM environment evolution is unitary, the reduced dynamics of the MWI alone is generically non-unitary.  
A natural and systematic framework to describe such dynamics is provided by the Schwinger-Keldysh (SK), or in--in, formalism~\cite{Keldysh:1964ud,Schwinger:1960qe}, which we use to construct an \textit{open} effective field theory (EFT) for a MWI.  
This formalism implements real-time evolution along a closed-time path (CTP), with forward ($+$) and backward ($-$) branches encoding the evolution of kets and bras, respectively, while preserving causality and unitarity at the level of the full system (for a modern review, see Ref.~\cite{Haehl2025}). 
Importantly, the degrees of freedom in the problem are doubled, with fields defined on the forward and backward time branches. 
This approach has been used in the context of interferometers to study gravitationally-induced decoherence~\cite{Blencowe:2012mp,Parikh:2020fhy,Kanno:2020usf}, soft-graviton~\cite{Zurek:2020ukz,Wilson-Gerow:2024ljx} and soft-photon~\cite{DeLisle:2022pjo} decoherence, and chameleon-induced effects~\cite{Burrage:2018pyg}, as well as the decay of the interference term between two Gaussian packets caused by a weakly coupled nonlinear bath of oscillators~\cite{Hu:1991di,Hu:1993vs}.

We find that phase and decoherence respond very differently to the statistics of the DM background. 
In the context of spin-independent elastic interactions, decoherence inherits novel Pauli-blocking and Bose-enhancement factors that are absent from the phase shift. 
The same formalism interpolates smoothly across the wave-particle boundary,  recovering collisional decoherence in the dilute heavy-probe limit.
We also capture the onset of non-Markovian dynamics when the DM coherence time exceeds the duration of the interferometric sequence. 
This is exemplified by the relation between contrast loss and the variance of the semiclassical phase shift due to ultralight linearly-coupled scalar DM.

\begin{figure}[t]
\centering
\begin{tikzpicture}[
  >=Stealth, line cap=round, line join=round, font=\footnotesize,
  x=0.84cm, y=0.84cm,
  axis/.style={line width=0.8pt, draw=black!65},
  tubeOuter/.style={draw=black!65, line width=32pt, line cap=butt},
  tubeInner/.style={draw=white, line width=31pt, line cap=butt},
  mainR/.style={line width=1.6pt, draw=orange!90!black},
  mainL/.style={line width=1.6pt, draw=green!55!black, dash pattern=on 4pt off 5pt}
]
\draw[axis,-{Stealth[length=2.2mm,width=1.4mm]}] (-4.0,-4.4) -- (-4.0,4.5);
\draw[axis,-{Stealth[length=2.2mm,width=1.4mm]}] (-4.0,-4.4) -- (4.4,-4.4);
\foreach \y/\lbl in {2.55/{$\frac{T+T'}{2}$},1.5/{$\frac{T}{2}$},0/{$0$},-1.5/{$-\frac{T}{2}$},-2.55/{$-\frac{T+T'}{2}$}}{
  \draw[axis] (-4.0,\y) -- ++(-0.13,0);
  \node[left=4pt, text=black!75] at (-4.18,\y) {\lbl};}
\foreach \x/\lbl in {-1.95/{$-\frac{\Delta x}{2}$},0/{$0$},1.95/{$\frac{\Delta x}{2}$}}{
  \draw[axis] (\x,-4.4) -- ++(0,-0.13);
  \node[below=3pt, text=black!75] at (\x,-4.55) {\lbl};}
\coordinate (Top) at (0,2.55); \coordinate (UR) at (1.95,1.5); \coordinate (LR) at (1.95,-1.5);
\coordinate (Bot) at (0,-2.55); \coordinate (LL) at (-1.95,-1.5); \coordinate (UL) at (-1.95,1.5);
\draw[tubeOuter] (0,4.2) -- (Top) -- (UR) -- (LR) -- (Bot) -- (0,-4.2);
\draw[tubeOuter] (0,4.2) -- (Top) -- (UL) -- (LL) -- (Bot) -- (0,-4.2);
\draw[tubeInner] (0,4.2) -- (Top) -- (UR) -- (LR) -- (Bot) -- (0,-4.2);
\draw[tubeInner] (0,4.2) -- (Top) -- (UL) -- (LL) -- (Bot) -- (0,-4.2);
\draw[orange!10!white, line width=31pt, line cap=butt] (0,4.2) -- (Top) -- (UR) -- (LR) -- (Bot) -- (0,-4.2);
\draw[orange!10!white, line width=31pt, line cap=butt] (0,4.2) -- (Top) -- (UL) -- (LL) -- (Bot) -- (0,-4.2);
\draw[mainR] (0,4.2) -- (Top) -- (UR) -- (LR) -- (Bot) -- (0,-4.2);
\draw[mainL] (0,4.2) -- (Top) -- (UL) -- (LL) -- (Bot) -- (0,-4.2);
\path (Top) -- (UL) coordinate[pos=0.54] (LabL);
\path (Top) -- (UR) coordinate[pos=0.54] (LabR);
\node[rotate=28.3, text=green!55!black] at ($(LabL)+(-0.15,0.28)$) {$\overline{\boldsymbol{x}}_L$};
\node[rotate=-28.3, text=orange!90!black] at ($(LabR)+(0.15,0.28)$) {$\overline{\boldsymbol{x}}_R$};
\draw[axis,<->,line width=0.6pt] (-0.63,-3.55) -- (0.63,-3.55);
\node[text=black!80] at (1.05,-3.4) {$\sigma$};
\end{tikzpicture}
\caption{Spacetime diagram for the MWI sequence considered in this work. The particular solutions of the unperturbed classical worldlines $\overline{\boldsymbol{x}}_{_{L,R}}(t)$ for a closed interferometric sequence geometry are shown in solid orange (right arm) and dashed green (left arm), together with the initial wavepacket of size $\sigma$ (shaded tube).}\label{fig:paths}
\end{figure}
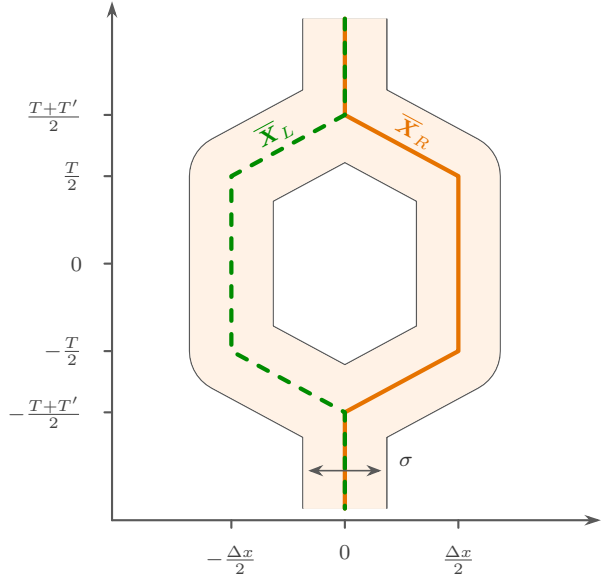

\textit{Open EFT for a single-atom MWI.}---For simplicity, we limit our attention to a single-atom MWI, and defer a comprehensive treatment of multi-atom platforms (e.g., atom interferometers) to Ref.~\cite{Badurina:2026myo}. 
At energies relevant for DM direct detection with matter-wave interferometry, atoms, although built from nucleons and electrons, can be treated as a single heavy degree of freedom of mass $M$. 
In a MWI, this degree of freedom is prepared in a superposition of states that are sent along different spacetime trajectories, as shown in Fig.~\ref{fig:paths}. 
In practice, this is achieved by entangling the c.o.m. degree of freedom with an internal degree of freedom, e.g., the atomic spin or energy level, which labels the trajectory. 
To make this structure manifest in the field theory, we introduce the path labels $\alpha, \beta \in\{L,R\}$ and describe the atom as a non-relativistic complex quantum field doublet $\Psi = (\psi_{_L},\psi_{_R})^T$ such that, upon canonical quantization, $\hat{\psi}_{\alpha}^\dagger(\mathbf{x})$ creates an atom at position $\mathbf{x}$ in path $\alpha$. 

In this work, we focus on monopole-monopole contact interactions between the system and the DM background populated by $\chi$ particles, which may be fermions or bosons.
Specifically, we consider the environment coupled to the atom through its charge density.
In the elastic regime, where the atom's internal state remains unchanged throughout the interferometric sequence, the internal structure is entirely encoded in a density profile $\varrho(\mathbf{r})$.
In our EFT, therefore, the interaction Lagrangian takes the form 
\begin{equation}\label{eq:Sint}
  \mathcal{L}_{\rm int}
  = \tfrac{g}{\Lambda^D}
    \int\!\mathrm{d}^3r\;
      \Psi^\dagger(t,\mathbf{x-r})\Psi(t,\mathbf{x-r})
      \,\varrho(\mathbf{r})
    \,O_\chi(x) \, ,
\end{equation}
where $g\ll1$ is a dimensionless coupling, $\Lambda$ is an energy scale, $D$ is chosen so that the mass dimension of the interaction term is four, $\mathbf{r}$ is the relative position of the charges, and $O_\chi$ is the model-dependent DM current, which depends on the DM fields $\chi$. 
In three-momentum space, the charge density yields the form factor $F_{A}(\mathbf{q})\equiv \int\!\mathrm{d}^3r \, e^{-\mi\mathbf{q}\cdot\mathbf{r}}\,\varrho(\mathbf{r})$.
We assume henceforth that the atom's charge $A$ under system-environment interactions is the same across both arms, and $\varrho(\mathbf{r})=A\,n(\mathbf{r})$, where $n(\mathbf{r})$ is the normalized spatial profile.

Since atoms are neither created nor destroyed at these energies, the dynamics and measurements are restricted to the one-particle sector of the Fock space of $\hat{\Psi}$. 
Indeed, for $\hat{\psi}_{\alpha}^\dagger(\mathbf{x})\ket{0} = \ket{\mathbf{x},\alpha}$, where $\ket{0}$ is the system's vacuum, the initial state of the system at initial time $t_i$ is simply
\begin{equation}\label{eq:initial_state}
\begin{aligned}
  \hat{\rho}_{_{s}}(t_i)
  &=
    \ket{\Phi}\!\bra{\Phi} \,, \\
  \ket{\Phi}
  &= \frac{1}{\sqrt{2}}\int\!\mathrm{d}^3x\;
    \Phi(\mathbf{x}) \bigl(\hat{\psi}_{L}^\dagger(\mathbf{x})+\hat{\psi}_{R}^\dagger(\mathbf{x})\bigr)\ket{0} \,,
\end{aligned}
\end{equation} 
where $\Phi(\mathbf{x})$ is the atomic wavepacket of width $\sigma$ centered at the origin.

The goal of this open EFT is to compute the reduced density matrix of the MWI at time $t_f$.
This is achieved by computing the joint unitary evolution of the apparatus (the system) and DM (the environment) from $t_i$ to $t_f$, given by the unitary operator $\hat{U}(t_f,t_i)$, after which the DM degrees of freedom are traced over, which we represent by $\mathrm{Tr}_e[\cdot]$. 
We assume a factorized initial state $\hat{\rho}_i=\hat{\rho}_{s}(t_i)\otimes\hat{\rho}_e$ at time $t_i$, with $\hat{\rho}_e$ the stationary state of the environment, and work in the weak-coupling (Born) approximation~\cite{Weiss:2021uhm}. 
After integrating over all space at time $t_f$, the system's reduced density matrix can be described as a two-by-two matrix in the $\{L,R\}$ basis:
\begin{widetext}
\begin{equation}
\label{eq:rho_PI}
\rho_s^{\alpha\beta}(t_f)  \equiv 
\!\! \int \!\! \mathrm{d}^3 x_f \bra{\alpha,\mathbf{x}_f}\mathrm{Tr}_e \left[\hat{U}(t_f,t_i) \,\hat{\rho}_i \,\hat{U}^\dagger(t_f,t_i)\right ]\ket{\beta,\mathbf{x}_f} \, ,
\end{equation}
\end{widetext}
where $\alpha, \beta \in \{L,R\}$. 
Importantly, the entries of this matrix can be written as a path integral over system worldlines $\boldsymbol{X}$~\cite{Sakurai:2011zz}. 
These worldlines are driven by external impulses parametrized by $\hat{J}$ with $\hat{J}\ket{\alpha}=J_\alpha \ket{\alpha}$, which models the beamsplitter and mirror pulses in atom interferometer experiments~\cite{Glick:2024xlq}.
Introducing the free system action 
\begin{equation}\label{eq:Ssys}
S_s[\boldsymbol{X}_\pm,\boldsymbol{J}_{\alpha,\pm}]\!=\!\!\int_{t_i}^{t_f}\!\mathrm{d}t'\Bigl[\tfrac{M}{2}|\dot{\boldsymbol{X}}_\pm|^2\!+\!\boldsymbol{J}_{\alpha,\pm}\!\cdot\!\boldsymbol{X}_\pm\Bigr]\, ,
\end{equation}
\begin{widetext}
Eq.~\eqref{eq:rho_PI} admits the worldline path integral representation 
\begin{equation}
\begin{aligned}\label{eq:rho_PI_worldline}
\rho_s^{\alpha\beta}(t_f) & \equiv 
\!\! \int\!\!\mathrm{d}^3 x_{i,+} \!\! \int \!\! \mathrm{d}^3 x_{i,-} \!\!\, \bra{\alpha,\mathbf{x}_{i,+}}\hat{\rho}_{_{s}}(t_i)\ket{\beta,\mathbf{x}_{i,-}} \\
& \qquad \times 
\!\! \int\!\!\mathrm{d}^3 x_f \int_{\mathbf{x}_{i,+}}^{\mathbf{x}_f}\!\!\!\!\mathcal{D}\boldsymbol{X}_+\!\!\int_{\mathbf{x}_{i,-}}^{\mathbf{x}_f}\!\!\!\!\mathcal{D}\boldsymbol{X}_-
e^{\mi S_s[\boldsymbol{X}_+,\boldsymbol{J}_{\alpha,+}]-\mi S_s[\boldsymbol{X}_-,\boldsymbol{J}_{\beta,-}]}\, 
e^{\mi S_{_\mathrm{IF}}[\boldsymbol{X}_+,\boldsymbol{X}_-]} \, , 
\end{aligned}
\end{equation}
with boundary conditions $\boldsymbol{X}_\pm(t_i)=\mathbf{x}_{i,\pm}$ and ${\boldsymbol{X}_\pm(t_f)=\mathbf{x}_f}$.
\end{widetext}

System-environment interactions and the initial state of the environment are captured by the Feynman--Vernon influence functional $e^{\mi S_{_\mathrm{IF}}[\boldsymbol{X}_+,\boldsymbol{X}_-]}$~\cite{Feynman:1963fq}, which we express in terms of the influence action $S_{_\mathrm{IF}}$. 
For concreteness, we assume that the environment is homogeneous and described by a boosted Maxwell--Boltzmann distribution $f_\chi(\mathbf{p})$, with $\int \tfrac{\mathrm{d}^3 p}{(2\pi)^3}f_\chi(\mathbf{p}) = 1$, as predicted by the Standard Halo Model in the laboratory frame~\cite{Drukier:1986tm}. 
We also assume that the DM state is Gaussian~\cite{Cheong:2024ose}. 
Assuming that each path of the interferometer couples in the same way to the environment, given Eq.~\eqref{eq:Sint}, expanding through $\mathcal{O}(g^2)$, and rotating to the Keldysh basis 
\begin{equation}
\begin{aligned}
O^{\text{cl}}(x) &= \tfrac{1}{2} \left [O_+(x) + O_-(x) \right ] \, ,\\
O^{\text{q}}(x) &= O_+(x) - O_-(x) \, ,
\end{aligned}
\end{equation}
the influence action takes the form 
\begin{widetext}
\begin{equation}\label{eq:Sif}
\begin{aligned}
S_{_{\rm IF}}[\boldsymbol{X}_+,\boldsymbol{X}_-] =& 
-\tfrac{1}{2}\!\left(\tfrac{g}{\Lambda^D}\right)^{\!2}\!\!
\int_{t_{i}}^{t_f}\!\! \mathrm{d}^4x\!\! 
\int_{t_{i}}^{t_f}\!\! \mathrm{d}^4x' \vec{\varrho}(x)^{\mathsf T}
\,\mathsf{K}(x - x')\,
\vec{\varrho}(x') + \mathcal{O}(g^3)\, ,
\end{aligned}
\end{equation}
\end{widetext}
with $\int_{t_{i}}^{t_f}\!\! \mathrm{d}^4x = \int_{t_{i}}^{t_f}\!\! \mathrm{d}t\int \mathrm{d}^3x$,
\begin{align}
\vec{\varrho}(x)^{\mathsf T} &=
\begin{pmatrix}
\varrho^{\mathrm{cl}}(x) & \varrho^{\mathrm{q}}(x)
\end{pmatrix} \label{eq:Os} \, , \\
\varrho^\mathrm{cl}(x)&=\tfrac{1}{2}\bigl[\varrho(\mathbf{x}\!-\!\boldsymbol{X}_+(t))+\varrho(\mathbf{x}\!-\!\boldsymbol{X}_-(t))\bigr]\, , \label{eq:rhocl} \\
\varrho^\mathrm{q}(x)&=\varrho(\mathbf{x}\!-\!\boldsymbol{X}_+(t))-\varrho(\mathbf{x}\!-\!\boldsymbol{X}_-(t))\,,\label{eq:rhoq}
\end{align}
and environment matrix kernel
\begin{align}
\mathsf{K}(x-x') &= \begin{pmatrix} 0 & K_{\text{a}}(x-x') \\ K_{\text{r}}(x-x') & K_{\text{k}}(x-x') \end{pmatrix},
\label{eq:Pi} \end{align}
defined in terms of the retarded, advanced and Keldysh kernels
\begin{align}
K_{\text{r}}(x-x') &= -\mi \langle\!\langle O^{\text{cl}}_\chi(x) O^{\text{q}}_\chi(x') \rangle\!\rangle_c \, , \label{eq:ret}\\
K_{\text{a}}(x-x') &= -\mi \langle\!\langle O^{\text{q}}_\chi(x) O^{\text{cl}}_\chi(x') \rangle\!\rangle_c \, ,\label{eq:adv}\\
K_{\text{k}}(x-x') &= -\mi \langle\!\langle O^{\text{cl}}_\chi(x) O^{\text{cl}}_\chi(x') \rangle\!\rangle_c \, , \label{eq:kel} 
\end{align}
respectively, 
where the $c$-subscript indicates connected correlation functions. 
Note that homogeneity implies that $\mathsf{K}$ depends exclusively on spacetime differences. 
The resulting integral kernels can therefore be nonlocal in time and space. 
Furthermore, because we assumed a closed interferometer geometry and both arms couple identically to the environment, the $\mathcal{O}(g)$ term in $S_\mathrm{IF}$ vanishes. 
Furthermore, the influence action manifestly satisfies the non-equilibrium constraint $S_{_\mathrm{IF}}[\boldsymbol{X}_+,\boldsymbol{X}_-]\!=\!0$ for $\boldsymbol{X}_+=\boldsymbol{X}_-$~\cite{Liu:2018kfw}.
Lastly, Gaussianity implies that $\mathsf{K}$ can be expressed in terms of two-point functions of $\chi$.

To evaluate the path integral, we decompose the worldlines into: (i) the classical unperturbed trajectories $\overline{\boldsymbol{X}}_\alpha$, which satisfy $M\ddot{\overline{\boldsymbol{X}}}_{{\alpha}}\!=\!\boldsymbol{J}_{{\alpha}}$ and are subject to the path integral's boundary conditions at $t_i$ and $t_f$, and (ii) worldline fluctuations that vanish at both endpoints. 
As we carefully explain in the Supplemental Material, the path integral can be organized as a series expansion in $1/M$. 
Corrections to the heavy-probe limit ($M\rightarrow \infty$ with $|\boldsymbol{J}_\alpha|/M$ fixed) are controlled by two dimensionless parameters:  $\mathbf{q}\cdot\mathbf{v}_\mathrm{rec} \tau_\mathrm{eff}$ and $|\mathbf{v}_\mathrm{rec}\tau_\mathrm{eff}|^2/\sigma^2$ where $\mathbf{v}_\mathrm{rec}=\mathbf{q}/M$ is the recoil velocity of the system after the environment transfers three-momentum $\mathbf{q}$ and $\tau_\mathrm{eff}=\min \left (T_s,\tau_e\right)$. $T_s$ is the duration of the experimental sequence ($T_s = T+T'$ in Fig.~\ref{fig:paths}) and $\tau_e$ is the time scale of the environment, e.g., the DM coherence time.
The first parameter encodes the phase accumulated as a result of a momentum kick $\mathbf{q}$ over time $\tau_\mathrm{eff}$, while the second measures the resolvability of the recoil against the wavepacket size.

In the $\{L,R\}$ basis and in the heavy-probe limit, the reduced density matrix takes the form 
\begin{equation}\label{eq:rho_heavy}
\begin{aligned}
\hat{\rho}_s(t_f) & \simeq \frac{1} {2}\begin{pmatrix} e^{\mi S_\mathrm{IF}[\overline{\boldsymbol{x}}_{L},\overline{\boldsymbol{x}}_{L}]} & e^{\mi S_\mathrm{IF}[\overline{\boldsymbol{x}}_{L},\overline{\boldsymbol{x}}_{R}]} \\ e^{\mi S_\mathrm{IF}[\overline{\boldsymbol{x}}_{R},\overline{\boldsymbol{x}}_{L}]} & e^{\mi  S_\mathrm{IF}[\overline{\boldsymbol{x}}_{R},\overline{\boldsymbol{x}}_{R}]} \end{pmatrix} \\
& \equiv  \frac{1}{2} \begin{pmatrix} 1 & e^{-\mathbb{D} - \mi \mathbb{P}} \\ e^{-\mathbb{D} +\mi \mathbb{P}} & 1 \end{pmatrix} \, ,
\end{aligned}
\end{equation}
with
\begin{align}
\mathbb{D}\!&=\!\mathrm{Im} \{  S_{_\mathrm{IF}}[\overline{\boldsymbol{x}}_L,\overline{\boldsymbol{x}}_R] \} \, , \\ 
\mathbb{P}\!&=-\mathrm{Re} \{ S_{_\mathrm{IF}}[\overline{\boldsymbol{x}}_L,\overline{\boldsymbol{x}}_R]  \} \, , 
\end{align}
the decoherence and phase, respectively, evaluated along the particular solutions $\overline{\boldsymbol{x}}_{\alpha}$ of the equations of motion of $\overline{\boldsymbol{X}}_\alpha$.

This result highlights a key difference between matter-wave interferometers and classical probes. 
Even in the absence of displacement recoils (i.e. $\boldsymbol{X}_\alpha(t) \rightarrow \overline{\boldsymbol{X}}_\alpha(t)$), a system prepared in a superposition of spatially separated states remains sensitive to environmental interactions. 
Indeed, at zeroth order in $1/M$, the environment changes the quantum correlations of the quantum system without modifying the system's paths. 
It is precisely this feature that allows MWIs to be exquisitely sensitive probes of DM via ultrasoft scattering processes.

\textit{Observables.}---We restrict our attention to the closed-interferometer geometry depicted in Fig.~\ref{fig:paths}, with maximal arm separation $|\boldsymbol{\Delta x}|$ held over the finite interrogation time $T\gg T'$.
Evaluating Eq.~\eqref{eq:Sif} on the classical (unperturbed) trajectories and taking the asymptotic time limit $t_f=-t_i = \mathcal{T} \rightarrow \infty$ yields
\begin{widetext}
\begin{align}
\mathbb{D}&=-T^2\!\left(\frac{g}{\Lambda^D}\right)^{\!2}\!\!\int\!\tfrac{\mathrm{d}^4 q}{(2\pi)^4}\;\mathrm{sinc}^2\!\Bigl(\tfrac{q_0T}{2}\Bigr)|F_A(\mathbf{q})|^2\bigl[1-\cos(\mathbf{q}\!\cdot\!\boldsymbol{\Delta x})\bigr]\mathrm{Im}\bigl[\widetilde{K}_\mathrm{k}(q)\bigr]\,,\label{eq:D_final}\\[4pt]
\mathbb{P}&=T\!\left(\frac{g}{\Lambda^D}\right)^{\!2}\!\!\int\!\tfrac{\mathrm{d}^4 q}{(2\pi)^4} \sin \Bigl(\tfrac{\mathbf{q} \cdot \boldsymbol{\Delta x}}{2}\Bigr)|F_A(\mathbf{q})|^2  \!\left[4\pi\delta(q_0)-4T\,\mathrm{sinc}^2\!\Bigl(\tfrac{q_0T}{2}\Bigr)\sin^2\!\Bigl(\tfrac{\mathbf{q} \cdot \boldsymbol{\Delta x}}{4}\Bigr)\right]\!\mathrm{Im}\bigl[\widetilde{K}_\mathrm{r}(q)\bigr]\,,\label{eq:P_final}
\end{align}
\end{widetext}
with $q$ the four momentum transferred from the environment to the system, as derived in the Supplemental Material. 

These expressions make the physical content of the open EFT transparent: \emph{decoherence is governed entirely by environmental fluctuations}, encoded in the noise kernel $K_\mathrm{k}$, while \emph{$\mathbb{P}$ is governed by causal influence}, captured by the retarded kernel $K_\mathrm{r}$. The latter vanishes at spacelike separation, enforcing microcausality. Physically, the classical MWI current induces a wake in the DM environment, whose coupling to the quantum MWI current produces the relative phase between the two arms. By contrast, $K_\mathrm{k}$ can remain nonzero at spacelike separations, allowing decoherence to persist without causal signal propagation~\cite{Calzetta:2008iqa}.

\textit{DM scattering signatures.}---We now apply our formalism to elastic spin-independent DM-nucleon scattering via a heavy hadrophilic scalar mediator $\phi$ with mass $m_\phi$.
After integrating out $\phi$, the microscopic interaction Lagrangian is
\begin{equation}\label{eq:Lscatt}
\begin{aligned}
\mathcal{L}_\mathrm{int} & \supset\frac{y_{_N} y_\chi}{m_\phi^2}\,\overline N N\,O_{\chi}\,, \\
O_{\chi}\!& =\!\begin{cases}\!
m_\chi\chi^2 & \text{for bosonic DM,} \\\!\overline{\chi}\chi & \text{for fermionic DM,}\end{cases}
\end{aligned}
\end{equation}
where $N$ denotes the nucleon, $\chi$ the DM, and $y_{_N}$ ($y_\chi$) the dimensionless coupling constant of the mediator to nucleons (DM) ~\cite{Knapen:2017xzo}.

For $T\!\gg\!\tau_c$, where $\tau_c \simeq 2\pi/m_\chi v_0^2$ is the coherence time, with $v_0 \simeq 10^{-3}$ the DM virial velocity, we recover the limit of Markovian dynamics. 
This limit is always satisfied for MWI searches of \textit{particle-like} sub-GeV DM~\cite{Riedel:2012ur,Riedel:2016acj,Du:2022ceh}.
In this case, one may invoke the distributional identity $\lim_{T\to\infty}T^2\mathrm{sinc}^2(q_0T/2)\!=\!2\pi T\,\delta(q_0)$, such that Eqs.~\eqref{eq:D_final}--\eqref{eq:P_final} may be rewritten as
\begin{align}
\mathbb{D}&=2\pi T\!\left(\tfrac{y_{_N} y_\chi}{m_\phi^2}\right)^{\!2}\!\!\int\!\tfrac{\mathrm{d}^3p}{(2\pi)^3} \!\!\int\! \tfrac{\mathrm{d}^3q}{(2\pi)^3}\, n_\chi(\mathbf{p})S_\chi(\mathbf{p,q})
\nonumber\\
&\quad\quad \times |F_A(\mathbf{q})|^2 [1-\cos(\mathbf{q}\!\cdot\!\boldsymbol{\Delta x})]\,\delta(E_\mathbf{p}\!-\!E_{\mathbf{p}-\mathbf{q}})\,,\label{eq:D_scatt}\\[2pt]
\mathbb{P}&=2\pi T\!\left(\tfrac{y_{_N} y_\chi}{m_\phi^2}\right)^{\!2}\!\!\int\!\tfrac{\mathrm{d}^3p}{(2\pi)^3}\!\!\int\!\tfrac{\mathrm{d}^3q}{(2\pi)^3} \,n_\chi(\mathbf{p}) \nonumber \\ &\quad\quad  \times|F_A(\mathbf{q})|^2
\sin(\mathbf{q}\!\cdot\!\boldsymbol{\Delta x})\,\delta(E_\mathbf{p}\!-\!E_{\mathbf{p}-\mathbf{q}})\,,\label{eq:P_scatt}
\end{align}
where $n_\chi(\mathbf{p}) = \tfrac{\rho_{_\mathrm{DM}}}{m_\chi} f_\chi(\mathbf{p})$ is the ({\em spin-summed}) DM occupation number per mode and $|F_A(\mathbf{q})|^2$ is the nuclear (or Helm) form factor~\cite{PhysRev.104.1466}. 
Here, we used the momentum representation of the environment kernels in the non-relativistic limit, as derived in the Supplemental Material, and introduced the DM statistical factor
\begin{equation}
\begin{aligned}
S_\chi(\mathbf{p,q}) &= \begin{cases}  1 + n_\chi(\mathbf{p-q}) \quad \text{for bosonic DM,} \\
1 - \tfrac{1}{2} n_\chi(\mathbf{p-q}) \quad \text{for fermionic DM.}
\end{cases}
\end{aligned}
\end{equation} 

These two expressions represent one of the central conceptual and phenomenological results of this work. 
In the dilute limit, $n_\chi\!\ll\!1$, our results reduce to the DM-induced phase and decoherence via collisional decoherence~\cite{Riedel:2012ur,Riedel:2016acj,Du:2022ceh}: decoherence saturates for $|\mathbf{q}|\!\gtrsim\! 1/|\boldsymbol{\Delta x}|$, and the phase shift can be nonzero for anisotropic environments~\cite{PhysRevA.42.38,Hornberger:2003umw}.
Beyond this limit, however, Eqs.~\eqref{eq:D_scatt}--\eqref{eq:P_scatt} highlight novel effects: from the statistical factor $S_\chi(\mathbf{p},\mathbf{q})$, $\mathbb{D}$ automatically acquires a \emph{(nonthermal) Bose enhancement factor} ${[1+ n_\chi(\mathbf{p}\!-\!\mathbf{q})]}$ for bosonic DM and \emph{Pauli blocking factor} $[1 - \tfrac{1}{2}n_\chi(\mathbf{p}\!-\!\mathbf{q})]$ for fermionic DM, for which $n_\chi(\mathbf{p})\leq2$. 
This behavior has \emph{no analog} in $\mathbb{P}$, which remains linear in the occupation number and does not acquire Bose-enhancement or Pauli-blocking factors.

The phenomenological consequences are striking. 
For fermionic DM, Pauli blocking drives $\mathbb{D}$ to zero as ${n_\chi(\mathbf{p})\to 2}$, while $\mathbb{P}$ remains finite. Remarkably, despite interacting with an environment, the MWI behaves effectively as a \textit{closed} quantum system. 
By contrast, for bosonic DM below the wave-particle boundary ($m_\chi\lesssim 10$~eV), the decoherence observable is enhanced relative to the phase shift by a factor ${n_\chi\simeq\lambda_\mathrm{c}^3\rho_{_\mathrm{DM}}/m_\chi\gg 1}$, where $\lambda_\mathrm{c}$ is the DM coherence length and $\rho_{_\mathrm{DM}}$ the local DM energy density. 
Therefore, depending on the assumptions regarding the DM mass and its spin statistics, one should target either phase- or decoherence-optimized readout schemes. 

\textit{Non-Markovian regime.}---When $T\!\lesssim\!\tau_c$, successive DM-MWI interactions remain correlated across the interferometric sequence. As a result, the environment retains memory over the interferometric sequence, and the sinc functions in Eqs.~\eqref{eq:D_final}--\eqref{eq:P_final} can no longer be approximated by delta functions. 
Since $E_\mathbf{p}-E_{\mathbf{p}-\mathbf{q}}\!     \lesssim\!\tau_c^{-1}$, in this limit $\mathbb{D}$ scales with $T^2$, while $\mathbb{P}$ contains terms that scale with $T^2$ and $T$. 
For typical interrogation times $T\sim\mathcal{O}(1\,\mathrm{s})$, the onset of non-Markovian behavior occurs for $m_\chi\!\lesssim\!10^{-9}$~eV. 
In this regime DM is ultralight, and the loss of contrast becomes the optimal observable. 

\emph{Linearly-coupled scalar DM}.---Finally, consider a real scalar field $\chi$ with $\mathcal{L}_\mathrm{int}\supset g\,\overline N N\,\chi$. 
In this case, the environment kernels coincide exactly with the DM Green's functions, $\widetilde{K}_{\mathrm{r,k}}\!=\!\widetilde{G}^\chi_{\mathrm{r,k}}$, the influence action reduces to the Feynman–Vernon form for a Gaussian environment~\cite{Feynman:1963fq}, and all $n_\chi$-dependent effects reside in $\widetilde{G}^\chi_\mathrm{k}$~\cite{Chou:1984es}. 
Therefore, $\mathbb{D}$ becomes the target observable. 

For $m_\chi T\gg1$, the energy-conserving Markovian contribution to $\mathbb{D}$ vanishes because the environment is gapped.
Physically, this occurs because the environment is gapped, no elastic interactions can occur, and therefore the environment cannot extract which-path information from the system. 
Hence, the MWI is again effectively a \textit{closed} quantum system, despite interacting with an environment of DM particles. 

In the limit $E_\mathbf{p} T\lesssim1$, finite-time effects smear energy conservation over $\sim 1/T$ and the $n_\chi(\mathbf{p})$-dependent part of the observable is now
\begin{equation}
\begin{aligned}
\mathbb{D}\!& =\!\tfrac{g^2 T^2}{m_\chi}\!\!\int\!\tfrac{\mathrm{d}^3p}{(2\pi)^3}\,\mathrm{sinc}^2\!\Bigl(\tfrac{E_\mathbf{p}T}{2}\Bigr) n_\chi(\mathbf{p}) \\
& \qquad \qquad \qquad \times |F_A(\mathbf{p})|^2 [1\!-\!\cos(\mathbf{p}\!\cdot\!\boldsymbol{\Delta x})] \, .
\end{aligned}
\end{equation}
In this case, decoherence can be interpreted as arising from the stimulated emission and absorption of $\chi$ particles with energies below the probe's energy scale. 
This is analogous to the process of decoherence due to bremsstrahlung of massless particles~\cite{PhysRevA.63.032102}. 
For typical interferometric times, this behavior occurs deep in the ultralight regime. In this regime, $\mathbb{D}$ is proportional to the variance of the semiclassical ultralight DM-induced phase ${\Delta \phi \simeq g A \int_{-T/2}^{T/2} dt \left [\chi(t,\mathbf{\Delta x}/2)-\chi(t,-\mathbf{\Delta x}/2)\right ]}$ \cite{Geraci:2016fva,Gue:2024onx} in the limit $T_\mathrm{exp}\gg \tau_c$, where $T_\mathrm{exp}$ is the total duration of the measurement campaign~\cite{Badurina:2022ngn}, and $A$ the atomic mass number.

\textit{Conclusions.}---Our open EFT, based on the SK formalism, highlights a structural asymmetry between the detectable phase and decoherence in MWIs. 
Decoherence ($\mathbb{D}$) is sourced by the environment's Keldysh kernel and inherits the full quantum statistics of the DM environment. The phase shift ($\mathbb{P}$) is sourced by the retarded kernel, and is insensitive to the full quantum statistics of the DM environment. 
For quadratically-coupled DM, $\mathbb{D}$ acquires Pauli-blocking and Bose-enhancement factors, while $\mathbb{P}$ remains at most linear in the DM occupation number. 
Phase and loss of contrast are therefore complementary observables, with their relative reach fixed by the DM mass, coupling structure, and spin statistics.

The framework further interpolates smoothly between the Markovian regime, in which the environment correlation time is short compared to the interferometric time, and the non-Markovian regime, where memory effects modify the scaling of both observables with the interferometric time. For example, in the case of ultralight linearly-coupled scalar DM, $\mathbb{D}$ is proportional to the variance of the semiclassical phase shift previously derived in the literature.

Finally, our results rigorously confirm that MWIs retain sensitivity to DM in the heavy-probe limit, in the absence of energy deposition. 
We expand upon these results and extend the same open-EFT machinery to multi-atom platforms in a companion paper~\cite{Badurina:2026myo}.

\begin{acknowledgments}
\textit{Acknowledgements.}---We thank Federico Cima, Roni Harnik, Junwu Huang, Michele Papucci and Allic Sivaramakrishnan for discussions, and Temple He for comments on the draft. 
L.B.\ and K.Z.\ are supported by the U.S.\ Department of Energy, Office of Science, Office of High Energy Physics under Award Number DE--SC0011632, and by the Walter Burke Institute for Theoretical Physics. 
K.Z.\ is also supported by a Simons Investigator award and by the Heising--Simons Foundation ``Observational Signatures of Quantum Gravity'' collaboration grant.
This work was performed in part at the Aspen Center for Physics, supported by NSF grant PHY--2210452, and the Mainz Institute for Theoretical Physics (MITP) of the Cluster of Excellence PRISMA\textsuperscript{+} (Project ID 390831469). GPT-6 Astra was used for proof-reading and to produce all figures. 
\end{acknowledgments}

\bibliography{bib}{}

\begin{thebibliography}{51}%
\makeatletter
\providecommand \@ifxundefined [1]{%
 \@ifx{#1\undefined}
}%
\providecommand \@ifnum [1]{%
 \ifnum #1\expandafter \@firstoftwo
 \else \expandafter \@secondoftwo
 \fi
}%
\providecommand \@ifx [1]{%
 \ifx #1\expandafter \@firstoftwo
 \else \expandafter \@secondoftwo
 \fi
}%
\providecommand \natexlab [1]{#1}%
\providecommand \enquote  [1]{``#1''}%
\providecommand \bibnamefont  [1]{#1}%
\providecommand \bibfnamefont [1]{#1}%
\providecommand \citenamefont [1]{#1}%
\providecommand \href@noop [0]{\@secondoftwo}%
\providecommand \href [0]{\begingroup \@sanitize@url \@href}%
\providecommand \@href[1]{\@@startlink{#1}\@@href}%
\providecommand \@@href[1]{\endgroup#1\@@endlink}%
\providecommand \@sanitize@url [0]{\catcode `\\12\catcode `\$12\catcode
  `\&12\catcode `\#12\catcode `\^12\catcode `\_12\catcode `\%12\relax}%
\providecommand \@@startlink[1]{}%
\providecommand \@@endlink[0]{}%
\providecommand \url  [0]{\begingroup\@sanitize@url \@url }%
\providecommand \@url [1]{\endgroup\@href {#1}{\urlprefix }}%
\providecommand \urlprefix  [0]{URL }%
\providecommand \Eprint [0]{\href }%
\providecommand \doibase [0]{https://doi.org/}%
\providecommand \selectlanguage [0]{\@gobble}%
\providecommand \bibinfo  [0]{\@secondoftwo}%
\providecommand \bibfield  [0]{\@secondoftwo}%
\providecommand \translation [1]{[#1]}%
\providecommand \BibitemOpen [0]{}%
\providecommand \bibitemStop [0]{}%
\providecommand \bibitemNoStop [0]{.\EOS\space}%
\providecommand \EOS [0]{\spacefactor3000\relax}%
\providecommand \BibitemShut  [1]{\csname bibitem#1\endcsname}%
\let\auto@bib@innerbib\@empty
\bibitem [{\citenamefont {Arvanitaki}\ \emph {et~al.}(2018)\citenamefont
  {Arvanitaki}, \citenamefont {Graham}, \citenamefont {Hogan}, \citenamefont
  {Rajendran},\ and\ \citenamefont {Van~Tilburg}}]{Arvanitaki:2016fyj}%
  \BibitemOpen
  \bibfield  {author} {\bibinfo {author} {\bibfnamefont {A.}~\bibnamefont
  {Arvanitaki}}, \bibinfo {author} {\bibfnamefont {P.~W.}\ \bibnamefont
  {Graham}}, \bibinfo {author} {\bibfnamefont {J.~M.}\ \bibnamefont {Hogan}},
  \bibinfo {author} {\bibfnamefont {S.}~\bibnamefont {Rajendran}},\ and\
  \bibinfo {author} {\bibfnamefont {K.}~\bibnamefont {Van~Tilburg}},\ }\href
  {https://doi.org/10.1103/PhysRevD.97.075020} {\bibfield  {journal} {\bibinfo
  {journal} {Phys. Rev. D}\ }\textbf {\bibinfo {volume} {97}},\ \bibinfo
  {pages} {075020} (\bibinfo {year} {2018})},\ \Eprint
  {https://arxiv.org/abs/1606.04541} {arXiv:1606.04541 [hep-ph]} \BibitemShut
  {NoStop}%
\bibitem [{\citenamefont {Badurina}\ \emph {et~al.}(2022)\citenamefont
  {Badurina}, \citenamefont {Blas},\ and\ \citenamefont
  {McCabe}}]{Badurina:2021lwr}%
  \BibitemOpen
  \bibfield  {author} {\bibinfo {author} {\bibfnamefont {L.}~\bibnamefont
  {Badurina}}, \bibinfo {author} {\bibfnamefont {D.}~\bibnamefont {Blas}},\
  and\ \bibinfo {author} {\bibfnamefont {C.}~\bibnamefont {McCabe}},\ }\href
  {https://doi.org/10.1103/PhysRevD.105.023006} {\bibfield  {journal} {\bibinfo
   {journal} {Phys. Rev. D}\ }\textbf {\bibinfo {volume} {105}},\ \bibinfo
  {pages} {023006} (\bibinfo {year} {2022})},\ \Eprint
  {https://arxiv.org/abs/2109.10965} {arXiv:2109.10965 [astro-ph.CO]}
  \BibitemShut {NoStop}%
\bibitem [{\citenamefont {Badurina}\ \emph
  {et~al.}(2023{\natexlab{a}})\citenamefont {Badurina}, \citenamefont
  {Beniwal},\ and\ \citenamefont {McCabe}}]{Badurina:2023wpk}%
  \BibitemOpen
  \bibfield  {author} {\bibinfo {author} {\bibfnamefont {L.}~\bibnamefont
  {Badurina}}, \bibinfo {author} {\bibfnamefont {A.}~\bibnamefont {Beniwal}},\
  and\ \bibinfo {author} {\bibfnamefont {C.}~\bibnamefont {McCabe}},\ }\href
  {https://doi.org/10.1103/PhysRevD.108.083016} {\bibfield  {journal} {\bibinfo
   {journal} {Phys. Rev. D}\ }\textbf {\bibinfo {volume} {108}},\ \bibinfo
  {pages} {083016} (\bibinfo {year} {2023}{\natexlab{a}})},\ \Eprint
  {https://arxiv.org/abs/2306.16477} {arXiv:2306.16477 [hep-ph]} \BibitemShut
  {NoStop}%
\bibitem [{\citenamefont {Badurina}\ \emph {et~al.}(2020)\citenamefont
  {Badurina} \emph {et~al.}}]{Badurina:2019hst}%
  \BibitemOpen
  \bibfield  {author} {\bibinfo {author} {\bibfnamefont {L.}~\bibnamefont
  {Badurina}} \emph {et~al.},\ }\href
  {https://doi.org/10.1088/1475-7516/2020/05/011} {\bibfield  {journal}
  {\bibinfo  {journal} {JCAP}\ }\textbf {\bibinfo {volume} {05}},\ \bibinfo
  {pages} {011}},\ \Eprint {https://arxiv.org/abs/1911.11755} {arXiv:1911.11755
  [astro-ph.CO]} \BibitemShut {NoStop}%
\bibitem [{\citenamefont {Banerjee}\ \emph {et~al.}(2023)\citenamefont
  {Banerjee}, \citenamefont {Perez}, \citenamefont {Safronova}, \citenamefont
  {Savoray},\ and\ \citenamefont {Shalit}}]{Banerjee:2022sqg}%
  \BibitemOpen
  \bibfield  {author} {\bibinfo {author} {\bibfnamefont {A.}~\bibnamefont
  {Banerjee}}, \bibinfo {author} {\bibfnamefont {G.}~\bibnamefont {Perez}},
  \bibinfo {author} {\bibfnamefont {M.}~\bibnamefont {Safronova}}, \bibinfo
  {author} {\bibfnamefont {I.}~\bibnamefont {Savoray}},\ and\ \bibinfo {author}
  {\bibfnamefont {A.}~\bibnamefont {Shalit}},\ }\href
  {https://doi.org/10.1007/JHEP10(2023)042} {\bibfield  {journal} {\bibinfo
  {journal} {JHEP}\ }\textbf {\bibinfo {volume} {10}},\ \bibinfo {pages}
  {042}},\ \Eprint {https://arxiv.org/abs/2211.05174} {arXiv:2211.05174
  [hep-ph]} \BibitemShut {NoStop}%
\bibitem [{\citenamefont {Blas}\ \emph {et~al.}(2025)\citenamefont {Blas},
  \citenamefont {Carlton},\ and\ \citenamefont {McCabe}}]{Blas:2024kps}%
  \BibitemOpen
  \bibfield  {author} {\bibinfo {author} {\bibfnamefont {D.}~\bibnamefont
  {Blas}}, \bibinfo {author} {\bibfnamefont {J.}~\bibnamefont {Carlton}},\ and\
  \bibinfo {author} {\bibfnamefont {C.}~\bibnamefont {McCabe}},\ }\href
  {https://doi.org/10.1103/zxtk-bwnf} {\bibfield  {journal} {\bibinfo
  {journal} {Phys. Rev. D}\ }\textbf {\bibinfo {volume} {111}},\ \bibinfo
  {pages} {115020} (\bibinfo {year} {2025})},\ \Eprint
  {https://arxiv.org/abs/2412.14282} {arXiv:2412.14282 [hep-ph]} \BibitemShut
  {NoStop}%
\bibitem [{\citenamefont {Beadle}\ \emph {et~al.}(2024)\citenamefont {Beadle},
  \citenamefont {Ellis}, \citenamefont {Quevillon},\ and\ \citenamefont
  {Hoa~Vuong}}]{Beadle:2023flm}%
  \BibitemOpen
  \bibfield  {author} {\bibinfo {author} {\bibfnamefont {C.}~\bibnamefont
  {Beadle}}, \bibinfo {author} {\bibfnamefont {S.~A.~R.}\ \bibnamefont
  {Ellis}}, \bibinfo {author} {\bibfnamefont {J.}~\bibnamefont {Quevillon}},\
  and\ \bibinfo {author} {\bibfnamefont {P.~N.}\ \bibnamefont {Hoa~Vuong}},\
  }\href {https://doi.org/10.1103/PhysRevD.110.035019} {\bibfield  {journal}
  {\bibinfo  {journal} {Phys. Rev. D}\ }\textbf {\bibinfo {volume} {110}},\
  \bibinfo {pages} {035019} (\bibinfo {year} {2024})},\ \Eprint
  {https://arxiv.org/abs/2307.10362} {arXiv:2307.10362 [hep-ph]} \BibitemShut
  {NoStop}%
\bibitem [{\citenamefont {Graham}\ \emph {et~al.}(2018)\citenamefont {Graham},
  \citenamefont {Kaplan}, \citenamefont {Mardon}, \citenamefont {Rajendran},
  \citenamefont {Terrano}, \citenamefont {Trahms},\ and\ \citenamefont
  {Wilkason}}]{Graham:2017ivz}%
  \BibitemOpen
  \bibfield  {author} {\bibinfo {author} {\bibfnamefont {P.~W.}\ \bibnamefont
  {Graham}}, \bibinfo {author} {\bibfnamefont {D.~E.}\ \bibnamefont {Kaplan}},
  \bibinfo {author} {\bibfnamefont {J.}~\bibnamefont {Mardon}}, \bibinfo
  {author} {\bibfnamefont {S.}~\bibnamefont {Rajendran}}, \bibinfo {author}
  {\bibfnamefont {W.~A.}\ \bibnamefont {Terrano}}, \bibinfo {author}
  {\bibfnamefont {L.}~\bibnamefont {Trahms}},\ and\ \bibinfo {author}
  {\bibfnamefont {T.}~\bibnamefont {Wilkason}},\ }\href
  {https://doi.org/10.1103/PhysRevD.97.055006} {\bibfield  {journal} {\bibinfo
  {journal} {Phys. Rev. D}\ }\textbf {\bibinfo {volume} {97}},\ \bibinfo
  {pages} {055006} (\bibinfo {year} {2018})},\ \Eprint
  {https://arxiv.org/abs/1709.07852} {arXiv:1709.07852 [hep-ph]} \BibitemShut
  {NoStop}%
\bibitem [{\citenamefont {Graham}\ \emph {et~al.}(2016)\citenamefont {Graham},
  \citenamefont {Kaplan}, \citenamefont {Mardon}, \citenamefont {Rajendran},\
  and\ \citenamefont {Terrano}}]{Graham:2015ifn}%
  \BibitemOpen
  \bibfield  {author} {\bibinfo {author} {\bibfnamefont {P.~W.}\ \bibnamefont
  {Graham}}, \bibinfo {author} {\bibfnamefont {D.~E.}\ \bibnamefont {Kaplan}},
  \bibinfo {author} {\bibfnamefont {J.}~\bibnamefont {Mardon}}, \bibinfo
  {author} {\bibfnamefont {S.}~\bibnamefont {Rajendran}},\ and\ \bibinfo
  {author} {\bibfnamefont {W.~A.}\ \bibnamefont {Terrano}},\ }\href
  {https://doi.org/10.1103/PhysRevD.93.075029} {\bibfield  {journal} {\bibinfo
  {journal} {Phys. Rev. D}\ }\textbf {\bibinfo {volume} {93}},\ \bibinfo
  {pages} {075029} (\bibinfo {year} {2016})},\ \Eprint
  {https://arxiv.org/abs/1512.06165} {arXiv:1512.06165 [hep-ph]} \BibitemShut
  {NoStop}%
\bibitem [{\citenamefont {Abe}\ \emph {et~al.}(2021)\citenamefont {Abe} \emph
  {et~al.}}]{MAGIS-100:2021etm}%
  \BibitemOpen
  \bibfield  {author} {\bibinfo {author} {\bibfnamefont {M.}~\bibnamefont
  {Abe}} \emph {et~al.} (\bibinfo {collaboration} {MAGIS-100}),\ }\href
  {https://doi.org/10.1088/2058-9565/abf719} {\bibfield  {journal} {\bibinfo
  {journal} {Quantum Sci. Technol.}\ }\textbf {\bibinfo {volume} {6}},\
  \bibinfo {pages} {044003} (\bibinfo {year} {2021})},\ \Eprint
  {https://arxiv.org/abs/2104.02835} {arXiv:2104.02835 [physics.atom-ph]}
  \BibitemShut {NoStop}%
\bibitem [{\citenamefont {Zhou}\ \emph {et~al.}(2024)\citenamefont {Zhou},
  \citenamefont {Ranson}, \citenamefont {Panagiotou},\ and\ \citenamefont
  {Overstreet}}]{Zhou:2024qdw}%
  \BibitemOpen
  \bibfield  {author} {\bibinfo {author} {\bibfnamefont {Y.}~\bibnamefont
  {Zhou}}, \bibinfo {author} {\bibfnamefont {R.}~\bibnamefont {Ranson}},
  \bibinfo {author} {\bibfnamefont {M.}~\bibnamefont {Panagiotou}},\ and\
  \bibinfo {author} {\bibfnamefont {C.}~\bibnamefont {Overstreet}},\ }\href
  {https://doi.org/10.1103/PhysRevA.110.033313} {\bibfield  {journal} {\bibinfo
   {journal} {Phys. Rev. A}\ }\textbf {\bibinfo {volume} {110}},\ \bibinfo
  {pages} {033313} (\bibinfo {year} {2024})},\ \Eprint
  {https://arxiv.org/abs/2406.00716} {arXiv:2406.00716 [physics.atom-ph]}
  \BibitemShut {NoStop}%
\bibitem [{\citenamefont {Riedel}(2013)}]{Riedel:2012ur}%
  \BibitemOpen
  \bibfield  {author} {\bibinfo {author} {\bibfnamefont {C.~J.}\ \bibnamefont
  {Riedel}},\ }\href {https://doi.org/10.1103/PhysRevD.88.116005} {\bibfield
  {journal} {\bibinfo  {journal} {Phys. Rev. D}\ }\textbf {\bibinfo {volume}
  {88}},\ \bibinfo {pages} {116005} (\bibinfo {year} {2013})},\ \Eprint
  {https://arxiv.org/abs/1212.3061} {arXiv:1212.3061 [quant-ph]} \BibitemShut
  {NoStop}%
\bibitem [{\citenamefont {Riedel}\ and\ \citenamefont
  {Yavin}(2017)}]{Riedel:2016acj}%
  \BibitemOpen
  \bibfield  {author} {\bibinfo {author} {\bibfnamefont {C.~J.}\ \bibnamefont
  {Riedel}}\ and\ \bibinfo {author} {\bibfnamefont {I.}~\bibnamefont {Yavin}},\
  }\href {https://doi.org/10.1103/PhysRevD.96.023007} {\bibfield  {journal}
  {\bibinfo  {journal} {Phys. Rev. D}\ }\textbf {\bibinfo {volume} {96}},\
  \bibinfo {pages} {023007} (\bibinfo {year} {2017})},\ \Eprint
  {https://arxiv.org/abs/1609.04145} {arXiv:1609.04145 [quant-ph]} \BibitemShut
  {NoStop}%
\bibitem [{\citenamefont {Du}\ \emph {et~al.}(2022)\citenamefont {Du},
  \citenamefont {Murgui}, \citenamefont {Pardo}, \citenamefont {Wang},\ and\
  \citenamefont {Zurek}}]{Du:2022ceh}%
  \BibitemOpen
  \bibfield  {author} {\bibinfo {author} {\bibfnamefont {Y.}~\bibnamefont
  {Du}}, \bibinfo {author} {\bibfnamefont {C.}~\bibnamefont {Murgui}}, \bibinfo
  {author} {\bibfnamefont {K.}~\bibnamefont {Pardo}}, \bibinfo {author}
  {\bibfnamefont {Y.}~\bibnamefont {Wang}},\ and\ \bibinfo {author}
  {\bibfnamefont {K.~M.}\ \bibnamefont {Zurek}},\ }\href
  {https://doi.org/10.1103/PhysRevD.106.095041} {\bibfield  {journal} {\bibinfo
   {journal} {Phys. Rev. D}\ }\textbf {\bibinfo {volume} {106}},\ \bibinfo
  {pages} {095041} (\bibinfo {year} {2022})},\ \Eprint
  {https://arxiv.org/abs/2205.13546} {arXiv:2205.13546 [hep-ph]} \BibitemShut
  {NoStop}%
\bibitem [{\citenamefont {Badurina}\ \emph {et~al.}(2024)\citenamefont
  {Badurina}, \citenamefont {Murgui},\ and\ \citenamefont
  {Plestid}}]{Badurina:2024nge}%
  \BibitemOpen
  \bibfield  {author} {\bibinfo {author} {\bibfnamefont {L.}~\bibnamefont
  {Badurina}}, \bibinfo {author} {\bibfnamefont {C.}~\bibnamefont {Murgui}},\
  and\ \bibinfo {author} {\bibfnamefont {R.}~\bibnamefont {Plestid}},\ }\href
  {https://doi.org/10.1103/PhysRevA.110.033311} {\bibfield  {journal} {\bibinfo
   {journal} {Phys. Rev. A}\ }\textbf {\bibinfo {volume} {110}},\ \bibinfo
  {pages} {033311} (\bibinfo {year} {2024})},\ \Eprint
  {https://arxiv.org/abs/2402.03421} {arXiv:2402.03421 [quant-ph]} \BibitemShut
  {NoStop}%
\bibitem [{\citenamefont {Badurina}\ \emph {et~al.}(2025)\citenamefont
  {Badurina}, \citenamefont {Du}, \citenamefont {Lee}, \citenamefont {Wang},\
  and\ \citenamefont {Zurek}}]{Badurina:2025xwl}%
  \BibitemOpen
  \bibfield  {author} {\bibinfo {author} {\bibfnamefont {L.}~\bibnamefont
  {Badurina}}, \bibinfo {author} {\bibfnamefont {Y.}~\bibnamefont {Du}},
  \bibinfo {author} {\bibfnamefont {V.~S.~H.}\ \bibnamefont {Lee}}, \bibinfo
  {author} {\bibfnamefont {Y.}~\bibnamefont {Wang}},\ and\ \bibinfo {author}
  {\bibfnamefont {K.~M.}\ \bibnamefont {Zurek}},\ }\href
  {https://doi.org/10.1103/xs7b-zgtj} {\bibfield  {journal} {\bibinfo
  {journal} {Phys. Rev. D}\ }\textbf {\bibinfo {volume} {112}},\ \bibinfo
  {pages} {063014} (\bibinfo {year} {2025})},\ \Eprint
  {https://arxiv.org/abs/2505.00781} {arXiv:2505.00781 [hep-ph]} \BibitemShut
  {NoStop}%
\bibitem [{\citenamefont {Badurina}\ \emph {et~al.}(2026)\citenamefont
  {Badurina}, \citenamefont {Blas}, \citenamefont {Ellis},\ and\ \citenamefont
  {Ellis}}]{Badurina:2025idj}%
  \BibitemOpen
  \bibfield  {author} {\bibinfo {author} {\bibfnamefont {L.}~\bibnamefont
  {Badurina}}, \bibinfo {author} {\bibfnamefont {D.}~\bibnamefont {Blas}},
  \bibinfo {author} {\bibfnamefont {J.}~\bibnamefont {Ellis}},\ and\ \bibinfo
  {author} {\bibfnamefont {S.~A.~R.}\ \bibnamefont {Ellis}},\ }\href
  {https://doi.org/10.1103/528c-xs6p} {\bibfield  {journal} {\bibinfo
  {journal} {Phys. Rev. D}\ }\textbf {\bibinfo {volume} {113}},\ \bibinfo
  {pages} {092004} (\bibinfo {year} {2026})},\ \Eprint
  {https://arxiv.org/abs/2507.17825} {arXiv:2507.17825 [hep-ph]} \BibitemShut
  {NoStop}%
\bibitem [{\citenamefont {Gallis}\ and\ \citenamefont
  {Fleming}(1990)}]{PhysRevA.42.38}%
  \BibitemOpen
  \bibfield  {author} {\bibinfo {author} {\bibfnamefont {M.~R.}\ \bibnamefont
  {Gallis}}\ and\ \bibinfo {author} {\bibfnamefont {G.~N.}\ \bibnamefont
  {Fleming}},\ }\href {https://doi.org/10.1103/PhysRevA.42.38} {\bibfield
  {journal} {\bibinfo  {journal} {Phys. Rev. A}\ }\textbf {\bibinfo {volume}
  {42}},\ \bibinfo {pages} {38} (\bibinfo {year} {1990})}\BibitemShut {NoStop}%
\bibitem [{\citenamefont {Hornberger}\ and\ \citenamefont
  {Sipe}(2003)}]{Hornberger:2003umw}%
  \BibitemOpen
  \bibfield  {author} {\bibinfo {author} {\bibfnamefont {K.}~\bibnamefont
  {Hornberger}}\ and\ \bibinfo {author} {\bibfnamefont {J.~E.}\ \bibnamefont
  {Sipe}},\ }\href {https://doi.org/10.1103/PhysRevA.68.012105} {\bibfield
  {journal} {\bibinfo  {journal} {Phys. Rev. A}\ }\textbf {\bibinfo {volume}
  {68}},\ \bibinfo {pages} {012105} (\bibinfo {year} {2003})},\ \Eprint
  {https://arxiv.org/abs/quant-ph/0303094} {arXiv:quant-ph/0303094}
  \BibitemShut {NoStop}%
\bibitem [{\citenamefont {Breuer}\ and\ \citenamefont
  {Petruccione}(2007)}]{Breuer:2007juk}%
  \BibitemOpen
  \bibfield  {author} {\bibinfo {author} {\bibfnamefont {H.-P.}\ \bibnamefont
  {Breuer}}\ and\ \bibinfo {author} {\bibfnamefont {F.}~\bibnamefont
  {Petruccione}},\ }\href
  {https://doi.org/10.1093/acprof:oso/9780199213900.001.0001} {\emph {\bibinfo
  {title} {{The Theory of Open Quantum Systems}}}}\ (\bibinfo  {publisher}
  {Oxford University Press},\ \bibinfo {year} {2007})\BibitemShut {NoStop}%
\bibitem [{\citenamefont {Keldysh}(1965)}]{Keldysh:1964ud}%
  \BibitemOpen
  \bibfield  {author} {\bibinfo {author} {\bibfnamefont {L.~V.}\ \bibnamefont
  {Keldysh}},\ }\href {https://doi.org/10.1142/9789811279461_0007} {\bibfield
  {journal} {\bibinfo  {journal} {Sov. Phys. JETP}\ }\textbf {\bibinfo {volume}
  {20}},\ \bibinfo {pages} {1018} (\bibinfo {year} {1965})}\BibitemShut
  {NoStop}%
\bibitem [{\citenamefont {Schwinger}(1961)}]{Schwinger:1960qe}%
  \BibitemOpen
  \bibfield  {author} {\bibinfo {author} {\bibfnamefont {J.~S.}\ \bibnamefont
  {Schwinger}},\ }\href {https://doi.org/10.1063/1.1703727} {\bibfield
  {journal} {\bibinfo  {journal} {J. Math. Phys.}\ }\textbf {\bibinfo {volume}
  {2}},\ \bibinfo {pages} {407} (\bibinfo {year} {1961})}\BibitemShut {NoStop}%
\bibitem [{\citenamefont {Haehl}\ and\ \citenamefont
  {Rangamani}(2025)}]{Haehl2025}%
  \BibitemOpen
  \bibfield  {author} {\bibinfo {author} {\bibfnamefont {F.}~\bibnamefont
  {Haehl}}\ and\ \bibinfo {author} {\bibfnamefont {M.}~\bibnamefont
  {Rangamani}},\ }\bibinfo {title} {Schwinger--keldysh formalism},\ in\ \href
  {https://doi.org/10.1007/978-3-031-90352-6_3} {\emph {\bibinfo {booktitle}
  {Records from the S-Matrix Marathon: Selected Topics on Scattering
  Amplitudes}}},\ \bibinfo {editor} {edited by\ \bibinfo {editor}
  {\bibfnamefont {N.}~\bibnamefont {Arkani-Hamed}}, \bibinfo {editor}
  {\bibfnamefont {M.}~\bibnamefont {Giroux}}, \bibinfo {editor} {\bibfnamefont
  {H.~S.}\ \bibnamefont {Hannesdottir}}, \bibinfo {editor} {\bibfnamefont
  {S.}~\bibnamefont {Mizera}},\ and\ \bibinfo {editor} {\bibfnamefont
  {C.}~\bibnamefont {Pasiecznik}}}\ (\bibinfo  {publisher} {Springer Nature
  Switzerland},\ \bibinfo {address} {Cham},\ \bibinfo {year} {2025})\ pp.\
  \bibinfo {pages} {89--129}\BibitemShut {NoStop}%
\bibitem [{\citenamefont {Blencowe}(2013)}]{Blencowe:2012mp}%
  \BibitemOpen
  \bibfield  {author} {\bibinfo {author} {\bibfnamefont {M.~P.}\ \bibnamefont
  {Blencowe}},\ }\href {https://doi.org/10.1103/PhysRevLett.111.021302}
  {\bibfield  {journal} {\bibinfo  {journal} {Phys. Rev. Lett.}\ }\textbf
  {\bibinfo {volume} {111}},\ \bibinfo {pages} {021302} (\bibinfo {year}
  {2013})},\ \Eprint {https://arxiv.org/abs/1211.4751} {arXiv:1211.4751
  [quant-ph]} \BibitemShut {NoStop}%
\bibitem [{\citenamefont {Parikh}\ \emph {et~al.}(2021)\citenamefont {Parikh},
  \citenamefont {Wilczek},\ and\ \citenamefont {Zahariade}}]{Parikh:2020fhy}%
  \BibitemOpen
  \bibfield  {author} {\bibinfo {author} {\bibfnamefont {M.}~\bibnamefont
  {Parikh}}, \bibinfo {author} {\bibfnamefont {F.}~\bibnamefont {Wilczek}},\
  and\ \bibinfo {author} {\bibfnamefont {G.}~\bibnamefont {Zahariade}},\ }\href
  {https://doi.org/10.1103/PhysRevD.104.046021} {\bibfield  {journal} {\bibinfo
   {journal} {Phys. Rev. D}\ }\textbf {\bibinfo {volume} {104}},\ \bibinfo
  {pages} {046021} (\bibinfo {year} {2021})},\ \Eprint
  {https://arxiv.org/abs/2010.08208} {arXiv:2010.08208 [hep-th]} \BibitemShut
  {NoStop}%
\bibitem [{\citenamefont {Kanno}\ \emph {et~al.}(2021)\citenamefont {Kanno},
  \citenamefont {Soda},\ and\ \citenamefont {Tokuda}}]{Kanno:2020usf}%
  \BibitemOpen
  \bibfield  {author} {\bibinfo {author} {\bibfnamefont {S.}~\bibnamefont
  {Kanno}}, \bibinfo {author} {\bibfnamefont {J.}~\bibnamefont {Soda}},\ and\
  \bibinfo {author} {\bibfnamefont {J.}~\bibnamefont {Tokuda}},\ }\href
  {https://doi.org/10.1103/PhysRevD.103.044017} {\bibfield  {journal} {\bibinfo
   {journal} {Phys. Rev. D}\ }\textbf {\bibinfo {volume} {103}},\ \bibinfo
  {pages} {044017} (\bibinfo {year} {2021})},\ \Eprint
  {https://arxiv.org/abs/2007.09838} {arXiv:2007.09838 [hep-th]} \BibitemShut
  {NoStop}%
\bibitem [{\citenamefont {Zurek}(2022)}]{Zurek:2020ukz}%
  \BibitemOpen
  \bibfield  {author} {\bibinfo {author} {\bibfnamefont {K.~M.}\ \bibnamefont
  {Zurek}},\ }\href {https://doi.org/10.1016/j.physletb.2022.136910} {\bibfield
   {journal} {\bibinfo  {journal} {Phys. Lett. B}\ }\textbf {\bibinfo {volume}
  {826}},\ \bibinfo {pages} {136910} (\bibinfo {year} {2022})},\ \Eprint
  {https://arxiv.org/abs/2012.05870} {arXiv:2012.05870 [hep-th]} \BibitemShut
  {NoStop}%
\bibitem [{\citenamefont {Wilson-Gerow}\ \emph {et~al.}(2024)\citenamefont
  {Wilson-Gerow}, \citenamefont {Dugad},\ and\ \citenamefont
  {Chen}}]{Wilson-Gerow:2024ljx}%
  \BibitemOpen
  \bibfield  {author} {\bibinfo {author} {\bibfnamefont {J.}~\bibnamefont
  {Wilson-Gerow}}, \bibinfo {author} {\bibfnamefont {A.}~\bibnamefont
  {Dugad}},\ and\ \bibinfo {author} {\bibfnamefont {Y.}~\bibnamefont {Chen}},\
  }\href {https://doi.org/10.1103/PhysRevD.110.045002} {\bibfield  {journal}
  {\bibinfo  {journal} {Phys. Rev. D}\ }\textbf {\bibinfo {volume} {110}},\
  \bibinfo {pages} {045002} (\bibinfo {year} {2024})},\ \Eprint
  {https://arxiv.org/abs/2405.00804} {arXiv:2405.00804 [hep-th]} \BibitemShut
  {NoStop}%
\bibitem [{\citenamefont {DeLisle}\ and\ \citenamefont
  {Stamp}(2024)}]{DeLisle:2022pjo}%
  \BibitemOpen
  \bibfield  {author} {\bibinfo {author} {\bibfnamefont {C.}~\bibnamefont
  {DeLisle}}\ and\ \bibinfo {author} {\bibfnamefont {P.~C.~E.}\ \bibnamefont
  {Stamp}},\ }\href {https://doi.org/10.1103/PhysRevA.110.022223} {\bibfield
  {journal} {\bibinfo  {journal} {Phys. Rev. A}\ }\textbf {\bibinfo {volume}
  {110}},\ \bibinfo {pages} {022223} (\bibinfo {year} {2024})},\ \Eprint
  {https://arxiv.org/abs/2211.05813} {arXiv:2211.05813 [quant-ph]} \BibitemShut
  {NoStop}%
\bibitem [{\citenamefont {Burrage}\ \emph {et~al.}(2019)\citenamefont
  {Burrage}, \citenamefont {K{\"a}ding}, \citenamefont {Millington},\ and\
  \citenamefont {Min{\'a}{\v{r}}}}]{Burrage:2018pyg}%
  \BibitemOpen
  \bibfield  {author} {\bibinfo {author} {\bibfnamefont {C.}~\bibnamefont
  {Burrage}}, \bibinfo {author} {\bibfnamefont {C.}~\bibnamefont {K{\"a}ding}},
  \bibinfo {author} {\bibfnamefont {P.}~\bibnamefont {Millington}},\ and\
  \bibinfo {author} {\bibfnamefont {J.}~\bibnamefont {Min{\'a}{\v{r}}}},\
  }\href {https://doi.org/10.1103/PhysRevD.100.076003} {\bibfield  {journal}
  {\bibinfo  {journal} {Phys. Rev. D}\ }\textbf {\bibinfo {volume} {100}},\
  \bibinfo {pages} {076003} (\bibinfo {year} {2019})},\ \Eprint
  {https://arxiv.org/abs/1812.08760} {arXiv:1812.08760 [hep-th]} \BibitemShut
  {NoStop}%
\bibitem [{\citenamefont {Hu}\ \emph {et~al.}(1992)\citenamefont {Hu},
  \citenamefont {Paz},\ and\ \citenamefont {Zhang}}]{Hu:1991di}%
  \BibitemOpen
  \bibfield  {author} {\bibinfo {author} {\bibfnamefont {B.~L.}\ \bibnamefont
  {Hu}}, \bibinfo {author} {\bibfnamefont {J.~P.}\ \bibnamefont {Paz}},\ and\
  \bibinfo {author} {\bibfnamefont {Y.-h.}\ \bibnamefont {Zhang}},\ }\href
  {https://doi.org/10.1103/PhysRevD.45.2843} {\bibfield  {journal} {\bibinfo
  {journal} {Phys. Rev. D}\ }\textbf {\bibinfo {volume} {45}},\ \bibinfo
  {pages} {2843} (\bibinfo {year} {1992})}\BibitemShut {NoStop}%
\bibitem [{\citenamefont {Hu}\ \emph {et~al.}(1993)\citenamefont {Hu},
  \citenamefont {Paz},\ and\ \citenamefont {Zhang}}]{Hu:1993vs}%
  \BibitemOpen
  \bibfield  {author} {\bibinfo {author} {\bibfnamefont {B.~L.}\ \bibnamefont
  {Hu}}, \bibinfo {author} {\bibfnamefont {J.~P.}\ \bibnamefont {Paz}},\ and\
  \bibinfo {author} {\bibfnamefont {Y.}~\bibnamefont {Zhang}},\ }\href
  {https://doi.org/10.1103/PhysRevD.47.1576} {\bibfield  {journal} {\bibinfo
  {journal} {Phys. Rev. D}\ }\textbf {\bibinfo {volume} {47}},\ \bibinfo
  {pages} {1576} (\bibinfo {year} {1993})}\BibitemShut {NoStop}%
\bibitem [{\citenamefont {Badurina}\ and\ \citenamefont
  {Zurek}(2026)}]{Badurina:2026myo}%
  \BibitemOpen
  \bibfield  {author} {\bibinfo {author} {\bibfnamefont {L.}~\bibnamefont
  {Badurina}}\ and\ \bibinfo {author} {\bibfnamefont {K.~M.}\ \bibnamefont
  {Zurek}},\ }\href@noop {} {\  (\bibinfo {year} {2026})},\ \Eprint
  {https://arxiv.org/abs/2610.36044} {arXiv:2610.36044 [hep-ph]} \BibitemShut
  {NoStop}%
\bibitem [{\citenamefont {Weiss}(2021)}]{Weiss:2021uhm}%
  \BibitemOpen
  \bibfield  {author} {\bibinfo {author} {\bibfnamefont {U.}~\bibnamefont
  {Weiss}},\ }\href {https://doi.org/10.1142/12402} {\emph {\bibinfo {title}
  {{Quantum Dissipative Systems}}}}\ (\bibinfo  {publisher} {World
  Scientific},\ \bibinfo {year} {2021})\BibitemShut {NoStop}%
\bibitem [{\citenamefont {Sakurai}\ and\ \citenamefont
  {Napolitano}(2020)}]{Sakurai:2011zz}%
  \BibitemOpen
  \bibfield  {author} {\bibinfo {author} {\bibfnamefont {J.~J.}\ \bibnamefont
  {Sakurai}}\ and\ \bibinfo {author} {\bibfnamefont {J.}~\bibnamefont
  {Napolitano}},\ }\href {https://doi.org/10.1017/9781108587280} {\emph
  {\bibinfo {title} {{Modern Quantum Mechanics}}}},\ \bibinfo {edition} {3rd}\
  ed.,\ Quantum physics, quantum information and quantum computation\ (\bibinfo
   {publisher} {Cambridge University Press},\ \bibinfo {year}
  {2020})\BibitemShut {NoStop}%
\bibitem [{\citenamefont {Glick}\ and\ \citenamefont
  {Kovachy}(2026)}]{Glick:2024xlq}%
  \BibitemOpen
  \bibfield  {author} {\bibinfo {author} {\bibfnamefont {J.}~\bibnamefont
  {Glick}}\ and\ \bibinfo {author} {\bibfnamefont {T.}~\bibnamefont
  {Kovachy}},\ }\href {https://doi.org/10.1116/5.0313350} {\bibfield  {journal}
  {\bibinfo  {journal} {AVS Quantum Sci.}\ }\textbf {\bibinfo {volume} {8}},\
  \bibinfo {pages} {024402} (\bibinfo {year} {2026})},\ \Eprint
  {https://arxiv.org/abs/2407.11446} {arXiv:2407.11446 [physics.atom-ph]}
  \BibitemShut {NoStop}%
\bibitem [{\citenamefont {Feynman}\ and\ \citenamefont
  {Vernon}(1963)}]{Feynman:1963fq}%
  \BibitemOpen
  \bibfield  {author} {\bibinfo {author} {\bibfnamefont {R.~P.}\ \bibnamefont
  {Feynman}}\ and\ \bibinfo {author} {\bibfnamefont {F.~L.}\ \bibnamefont
  {Vernon}, \bibfnamefont {Jr.}},\ }\href
  {https://doi.org/10.1016/0003-4916(63)90068-X} {\bibfield  {journal}
  {\bibinfo  {journal} {Annals Phys.}\ }\textbf {\bibinfo {volume} {24}},\
  \bibinfo {pages} {118} (\bibinfo {year} {1963})}\BibitemShut {NoStop}%
\bibitem [{\citenamefont {Drukier}\ \emph {et~al.}(1986)\citenamefont
  {Drukier}, \citenamefont {Freese},\ and\ \citenamefont
  {Spergel}}]{Drukier:1986tm}%
  \BibitemOpen
  \bibfield  {author} {\bibinfo {author} {\bibfnamefont {A.~K.}\ \bibnamefont
  {Drukier}}, \bibinfo {author} {\bibfnamefont {K.}~\bibnamefont {Freese}},\
  and\ \bibinfo {author} {\bibfnamefont {D.~N.}\ \bibnamefont {Spergel}},\
  }\href {https://doi.org/10.1103/PhysRevD.33.3495} {\bibfield  {journal}
  {\bibinfo  {journal} {Phys. Rev. D}\ }\textbf {\bibinfo {volume} {33}},\
  \bibinfo {pages} {3495} (\bibinfo {year} {1986})}\BibitemShut {NoStop}%
\bibitem [{\citenamefont {Cheong}\ \emph {et~al.}(2025)\citenamefont {Cheong},
  \citenamefont {Rodd},\ and\ \citenamefont {Wang}}]{Cheong:2024ose}%
  \BibitemOpen
  \bibfield  {author} {\bibinfo {author} {\bibfnamefont {D.~Y.}\ \bibnamefont
  {Cheong}}, \bibinfo {author} {\bibfnamefont {N.~L.}\ \bibnamefont {Rodd}},\
  and\ \bibinfo {author} {\bibfnamefont {L.-T.}\ \bibnamefont {Wang}},\ }\href
  {https://doi.org/10.1103/PhysRevD.111.015028} {\bibfield  {journal} {\bibinfo
   {journal} {Phys. Rev. D}\ }\textbf {\bibinfo {volume} {111}},\ \bibinfo
  {pages} {015028} (\bibinfo {year} {2025})},\ \Eprint
  {https://arxiv.org/abs/2408.04696} {arXiv:2408.04696 [hep-ph]} \BibitemShut
  {NoStop}%
\bibitem [{\citenamefont {Liu}\ and\ \citenamefont
  {Glorioso}(2018)}]{Liu:2018kfw}%
  \BibitemOpen
  \bibfield  {author} {\bibinfo {author} {\bibfnamefont {H.}~\bibnamefont
  {Liu}}\ and\ \bibinfo {author} {\bibfnamefont {P.}~\bibnamefont {Glorioso}},\
  }\href {https://doi.org/10.22323/1.305.0008} {\bibfield  {journal} {\bibinfo
  {journal} {PoS}\ }\textbf {\bibinfo {volume} {TASI2017}},\ \bibinfo {pages}
  {008} (\bibinfo {year} {2018})},\ \Eprint {https://arxiv.org/abs/1805.09331}
  {arXiv:1805.09331 [hep-th]} \BibitemShut {NoStop}%
\bibitem [{\citenamefont {Calzetta}\ and\ \citenamefont
  {Hu}(2009)}]{Calzetta:2008iqa}%
  \BibitemOpen
  \bibfield  {author} {\bibinfo {author} {\bibfnamefont {E.~A.}\ \bibnamefont
  {Calzetta}}\ and\ \bibinfo {author} {\bibfnamefont {B.-L.~B.}\ \bibnamefont
  {Hu}},\ }\href {https://doi.org/10.1017/9781009290036} {\emph {\bibinfo
  {title} {{Nonequilibrium Quantum Field Theory}}}}\ (\bibinfo  {publisher}
  {Oxford University Press},\ \bibinfo {year} {2009})\BibitemShut {NoStop}%
\bibitem [{\citenamefont {Knapen}\ \emph {et~al.}(2017)\citenamefont {Knapen},
  \citenamefont {Lin},\ and\ \citenamefont {Zurek}}]{Knapen:2017xzo}%
  \BibitemOpen
  \bibfield  {author} {\bibinfo {author} {\bibfnamefont {S.}~\bibnamefont
  {Knapen}}, \bibinfo {author} {\bibfnamefont {T.}~\bibnamefont {Lin}},\ and\
  \bibinfo {author} {\bibfnamefont {K.~M.}\ \bibnamefont {Zurek}},\ }\href
  {https://doi.org/10.1103/PhysRevD.96.115021} {\bibfield  {journal} {\bibinfo
  {journal} {Phys. Rev. D}\ }\textbf {\bibinfo {volume} {96}},\ \bibinfo
  {pages} {115021} (\bibinfo {year} {2017})},\ \Eprint
  {https://arxiv.org/abs/1709.07882} {arXiv:1709.07882 [hep-ph]} \BibitemShut
  {NoStop}%
\bibitem [{\citenamefont {Helm}(1956)}]{PhysRev.104.1466}%
  \BibitemOpen
  \bibfield  {author} {\bibinfo {author} {\bibfnamefont {R.~H.}\ \bibnamefont
  {Helm}},\ }\href {https://doi.org/10.1103/PhysRev.104.1466} {\bibfield
  {journal} {\bibinfo  {journal} {Phys. Rev.}\ }\textbf {\bibinfo {volume}
  {104}},\ \bibinfo {pages} {1466} (\bibinfo {year} {1956})}\BibitemShut
  {NoStop}%
\bibitem [{\citenamefont {Chou}\ \emph {et~al.}(1985)\citenamefont {Chou},
  \citenamefont {Su}, \citenamefont {Hao},\ and\ \citenamefont
  {Yu}}]{Chou:1984es}%
  \BibitemOpen
  \bibfield  {author} {\bibinfo {author} {\bibfnamefont {K.-c.}\ \bibnamefont
  {Chou}}, \bibinfo {author} {\bibfnamefont {Z.-b.}\ \bibnamefont {Su}},
  \bibinfo {author} {\bibfnamefont {B.-l.}\ \bibnamefont {Hao}},\ and\ \bibinfo
  {author} {\bibfnamefont {L.}~\bibnamefont {Yu}},\ }\href
  {https://doi.org/10.1016/0370-1573(85)90136-X} {\bibfield  {journal}
  {\bibinfo  {journal} {Phys. Rept.}\ }\textbf {\bibinfo {volume} {118}},\
  \bibinfo {pages} {1} (\bibinfo {year} {1985})}\BibitemShut {NoStop}%
\bibitem [{\citenamefont {Breuer}\ and\ \citenamefont
  {Petruccione}(2001)}]{PhysRevA.63.032102}%
  \BibitemOpen
  \bibfield  {author} {\bibinfo {author} {\bibfnamefont {H.-P.}\ \bibnamefont
  {Breuer}}\ and\ \bibinfo {author} {\bibfnamefont {F.}~\bibnamefont
  {Petruccione}},\ }\href {https://doi.org/10.1103/PhysRevA.63.032102}
  {\bibfield  {journal} {\bibinfo  {journal} {Phys. Rev. A}\ }\textbf {\bibinfo
  {volume} {63}},\ \bibinfo {pages} {032102} (\bibinfo {year}
  {2001})}\BibitemShut {NoStop}%
\bibitem [{\citenamefont {Geraci}\ and\ \citenamefont
  {Derevianko}(2016)}]{Geraci:2016fva}%
  \BibitemOpen
  \bibfield  {author} {\bibinfo {author} {\bibfnamefont {A.~A.}\ \bibnamefont
  {Geraci}}\ and\ \bibinfo {author} {\bibfnamefont {A.}~\bibnamefont
  {Derevianko}},\ }\href {https://doi.org/10.1103/PhysRevLett.117.261301}
  {\bibfield  {journal} {\bibinfo  {journal} {Phys. Rev. Lett.}\ }\textbf
  {\bibinfo {volume} {117}},\ \bibinfo {pages} {261301} (\bibinfo {year}
  {2016})},\ \Eprint {https://arxiv.org/abs/1605.04048} {arXiv:1605.04048
  [physics.atom-ph]} \BibitemShut {NoStop}%
\bibitem [{\citenamefont {Gu{\'e}}\ \emph {et~al.}(2024)\citenamefont
  {Gu{\'e}}, \citenamefont {Hees},\ and\ \citenamefont {Wolf}}]{Gue:2024onx}%
  \BibitemOpen
  \bibfield  {author} {\bibinfo {author} {\bibfnamefont {J.}~\bibnamefont
  {Gu{\'e}}}, \bibinfo {author} {\bibfnamefont {A.}~\bibnamefont {Hees}},\ and\
  \bibinfo {author} {\bibfnamefont {P.}~\bibnamefont {Wolf}},\ }\href
  {https://doi.org/10.1103/PhysRevD.110.035005} {\bibfield  {journal} {\bibinfo
   {journal} {Phys. Rev. D}\ }\textbf {\bibinfo {volume} {110}},\ \bibinfo
  {pages} {035005} (\bibinfo {year} {2024})},\ \Eprint
  {https://arxiv.org/abs/2401.14742} {arXiv:2401.14742 [hep-ph]} \BibitemShut
  {NoStop}%
\bibitem [{\citenamefont {Badurina}\ \emph
  {et~al.}(2023{\natexlab{b}})\citenamefont {Badurina}, \citenamefont {Gibson},
  \citenamefont {McCabe},\ and\ \citenamefont {Mitchell}}]{Badurina:2022ngn}%
  \BibitemOpen
  \bibfield  {author} {\bibinfo {author} {\bibfnamefont {L.}~\bibnamefont
  {Badurina}}, \bibinfo {author} {\bibfnamefont {V.}~\bibnamefont {Gibson}},
  \bibinfo {author} {\bibfnamefont {C.}~\bibnamefont {McCabe}},\ and\ \bibinfo
  {author} {\bibfnamefont {J.}~\bibnamefont {Mitchell}},\ }\href
  {https://doi.org/10.1103/PhysRevD.107.055002} {\bibfield  {journal} {\bibinfo
   {journal} {Phys. Rev. D}\ }\textbf {\bibinfo {volume} {107}},\ \bibinfo
  {pages} {055002} (\bibinfo {year} {2023}{\natexlab{b}})},\ \Eprint
  {https://arxiv.org/abs/2211.01854} {arXiv:2211.01854 [hep-ph]} \BibitemShut
  {NoStop}%
\bibitem [{\citenamefont {Corradini}\ \emph {et~al.}(2015)\citenamefont
  {Corradini}, \citenamefont {Schubert}, \citenamefont {Edwards},\ and\
  \citenamefont {Ahmadiniaz}}]{Corradini:2015tik}%
  \BibitemOpen
  \bibfield  {author} {\bibinfo {author} {\bibfnamefont {O.}~\bibnamefont
  {Corradini}}, \bibinfo {author} {\bibfnamefont {C.}~\bibnamefont {Schubert}},
  \bibinfo {author} {\bibfnamefont {J.~P.}\ \bibnamefont {Edwards}},\ and\
  \bibinfo {author} {\bibfnamefont {N.}~\bibnamefont {Ahmadiniaz}}\ }(\bibinfo
  {year} {2015})\ \Eprint {https://arxiv.org/abs/1512.08694} {arXiv:1512.08694
  [hep-th]} \BibitemShut {NoStop}%
\bibitem [{\citenamefont {Kamenev}(2011)}]{Kamenev_2011}%
  \BibitemOpen
  \bibfield  {author} {\bibinfo {author} {\bibfnamefont {A.}~\bibnamefont
  {Kamenev}},\ }\href@noop {} {\emph {\bibinfo {title} {Field Theory of
  Non-Equilibrium Systems}}}\ (\bibinfo  {publisher} {Cambridge University
  Press},\ \bibinfo {year} {2011})\BibitemShut {NoStop}%
\bibitem [{\citenamefont {Larkin}\ and\ \citenamefont
  {Ovchinnikov}(1975)}]{Larkin_Ovchinnikov_1975}%
  \BibitemOpen
  \bibfield  {author} {\bibinfo {author} {\bibfnamefont {A.~I.}\ \bibnamefont
  {Larkin}}\ and\ \bibinfo {author} {\bibfnamefont {Y.~N.}\ \bibnamefont
  {Ovchinnikov}},\ }\href {http://jetp.ras.ru/cgi-bin/dn/e_041_05_0960.pdf}
  {\bibfield  {journal} {\bibinfo  {journal} {Journal of Experimental and
  Theoretical Physics}\ }\textbf {\bibinfo {volume} {41}},\ \bibinfo {pages}
  {960} (\bibinfo {year} {1975})},\ \bibinfo {note} {russian original: Zh.
  Eksp. Teor. Fiz. 68, 1915-1927 (1975)}\BibitemShut {NoStop}%
\end{thebibliography}%
\bibliographystyle{apsrev4-2}

\clearpage
\widetext

\setcounter{equation}{0}
\setcounter{figure}{0}
\setcounter{table}{0}
\setcounter{section}{0}

\renewcommand{\theequation}{S\arabic{equation}}
\renewcommand{\thefigure}{S\arabic{figure}}
\renewcommand{\thetable}{S\arabic{table}}
\renewcommand*{\thesection}{S.\Roman{section}}

\begin{center}
\textbf{\large Matter-wave Interferometers as Open-system Dark Matter Detectors} \\
\vspace{0.2cm}
\textit{ \large Supplemental Material} \\
\vspace{0.2cm}
Leonardo Badurina and Kathryn M. Zurek
\end{center}
\vspace{0.2cm}

This Supplemental Material collects the technical results that underlie the main text. 
We use the conventions $x_\mu x^\mu=t^2-|\mathbf{x}|^2$ $\int\dbar^n p=\int\mathrm{d}^n p/(2\pi)^n$, $\deltabar^{(n)}(\cdot)=(2\pi)^n\delta^{(n)}(\cdot)$, and natural units throughout. 
The four-momentum Fourier transform (FT) of a function $f(x)$ and its inverse FT are respectively defined as
\begin{equation}\label{eq:FT}
\begin{aligned}
\widetilde{f}(p) = \int \mathrm{d}^4 x \, f(x) e^{\mi p\cdot x} \, \quad , \quad f(x)= \int \dbar^4 p \, \widetilde{f}(p) e^{-\mi p\cdot x} \, ,
\end{aligned}
\end{equation}
while the three-momentum FT and its inverse are
\begin{equation}\label{eq:FT_3}
\begin{aligned}
\widetilde{f}(\mathbf{p},t) = \int \mathrm{d}^3 x \, f(x) e^{-\mi \mathbf{p}\cdot \mathbf{x}} \, \quad , \quad f(x)= \int \dbar^3 p \, \widetilde{f}(\mathbf{p},t) e^{\mi \mathbf{p}\cdot \mathbf{x}} \, .
\end{aligned}
\end{equation}

\section{Heavy-probe limit, asymptotics and the system reduced density matrix}\label{sec:heavy}

In this section, we outline how to obtain the heavy-probe limit of the system's reduced density matrix. We begin by decomposing the system worldlines $\boldsymbol{X}_\pm(t)$ as
\begin{equation}\label{eq:decomp}
\boldsymbol{X}_\pm(t) = \overline{\boldsymbol{X}}_{\alpha,\pm}(t)+\boldsymbol{\delta X}_\pm(t) \, ,
\end{equation}
where $\overline{\boldsymbol{X}}_{\alpha,\pm}(t)$ solves the classical equations of motion in the presence of the physical external currents $\boldsymbol{J}_{\alpha,\pm}(t)$ and satisfies the path integral's boundary conditions, i.e.
\begin{equation}\label{eq:unperturbed}
M \ddot{\overline{\boldsymbol{X}}}_{\alpha,\pm} = \boldsymbol{J}_{\alpha,\pm} \,, \qquad 
\overline{\boldsymbol{X}}_{\alpha,\pm}(t_i) = \mathbf{x}_{i,\pm} \,, \qquad 
\overline{\boldsymbol{X}}_{\alpha,\pm}(t_f) =  \mathbf{x}_f \,, 
\end{equation}
and $\boldsymbol{\delta X}_\pm$ captures worldline fluctuations obeying the Dirichlet boundary conditions
\begin{equation}\label{eq:dX_BC}
\boldsymbol{\delta X}_\pm(t_i) = \boldsymbol{\delta X}_\pm(t_f) = \mathbf{0} \,.
\end{equation}
Under this decomposition, the free system action along either time branch may be expressed as
\begin{equation}
\begin{aligned}
S_s[\overline{\boldsymbol{X}}_{\alpha, \pm}+\boldsymbol{\delta X}_\pm,\boldsymbol{J}_{\alpha,\pm}] 
&= S_s[\overline{\boldsymbol{X}}_{\alpha,\pm},\boldsymbol{J}_{\alpha,\pm}] + S_s[\delta \boldsymbol{X}_{\pm}] \, ,
\end{aligned}
\end{equation}
where 
\begin{equation}\label{app:eq:Scl}
\begin{aligned}
S_s[\overline{\boldsymbol{X}}_{\alpha, \pm},\boldsymbol{J}_{\alpha, \pm}]
= \tfrac{M}{2}\left [ \dot{\overline{\boldsymbol{X}}}_{\alpha, \pm} \cdot \overline{\boldsymbol{X}}_{\alpha, \pm} \right ]^{t_f}_{t_i} + \tfrac{1}{2}\int_{t_i}^{t_f}\!\mathrm{d}t \boldsymbol{J}_{\alpha, \pm}\!\cdot\!\overline{\boldsymbol{X}}_{\alpha, \pm}  \, ,
\end{aligned}
\end{equation}
is the free classical action, and
\begin{equation}
\begin{aligned}
S_s[\boldsymbol{\delta X}_\pm]\!&=\!\!\int_{t_i}^{t_f}\!\mathrm{d}t \tfrac{M}{2}|\dot{\boldsymbol{\delta X}}_\pm|^2 \, ,
\end{aligned}
\end{equation}
is the free action for worldline fluctuations. 
Hence, the ($\alpha,\beta$)-entry of the system's reduced density matrix (cf. Eq.~\eqref{eq:rho_PI_worldline}) can be rewritten as
\begin{equation}\label{app:eq:rhos}
\begin{aligned}
\rho_s^{\alpha \beta}(t_f) & =\!\int\!\!\mathrm{d}^3 x_{i,+}\int \mathrm{d}^3 x_{i,-} \!\!\, \bra{\alpha,\mathbf{x}_{i,+}}\hat{\rho}_{_{s}}(t_i)\ket{\beta,\mathbf{x}_{i,-}} 
\int\!\!\mathrm{d}^3 x_f \,  e^{\mi S_s[\overline{\boldsymbol{X}}_{\alpha,+},\boldsymbol{J}_{\alpha,+}]-\mi S_s[\overline{\boldsymbol{X}}_{\beta,-},\boldsymbol{J}_{\beta,-}]} \\
 & \qquad \qquad \qquad \qquad \times \int_{\mathbf{0}}^{\mathbf{0}}\!\!\!\!\mathcal{D}\boldsymbol{\delta X}_+\!\!\int_{\mathbf{0}}^{\mathbf{0}}\!\!\mathcal{D}\boldsymbol{\delta X}_-
e^{\mi S_s[\delta \boldsymbol{X}_+]-\mi S_s[\delta \boldsymbol{X}_-]}\, 
e^{\mi S_{_\mathrm{IF}}[\overline{\boldsymbol{X}}_{\alpha,+}+\boldsymbol{\delta X}_+,\overline{\boldsymbol{X}}_{\beta,-}+\boldsymbol{\delta X}_-]}  \, . 
\end{aligned}
\end{equation}
Expanding around the classical worldlines corresponds to expressing the path integral as a power series expansion in $\delta X_\pm$. 
Because of the structure of the path integral over fluctuations, it is advantageous to: (i) introduce fictitious currents $\boldsymbol{j}_\pm$ that couple linearly to the worldline fluctuations; (ii) rewrite $\boldsymbol{\delta X_\pm} \rightarrow \mp \mi \boldsymbol{\delta}/\boldsymbol{\delta j}_\pm$ in the influence action, in order to pull $S_\mathrm{IF}$ out of the path integral over fluctuations; (iii) perform the exact Gaussian path integral for fluctuations; and (iv) compute corrections to the reduced density matrix due to worldline fluctuations by linearizing the influence functional, acting with the functional derivatives in $S_\mathrm{IF}$ on the free worldline path integral, and then setting $\boldsymbol{j}_\pm = 0$, i.e.,
\begin{equation}\label{app:eq:step2}
\begin{aligned}
\text{2\textsuperscript{nd} line of Eq.~\eqref{app:eq:rhos}}  & = e^{\mi S_{_\mathrm{IF}}\left [\overline{\boldsymbol{X}}_{\alpha,+} - i \tfrac{\boldsymbol{\delta}}{\boldsymbol{\delta j}_+},\overline{\boldsymbol{X}}_{\beta,-}+ i \tfrac{\boldsymbol{\delta}}{\boldsymbol{\delta j}_-}\right ]}  \int_{\mathbf{0}}^{\mathbf{0}}\!\!\!\!\mathcal{D}\boldsymbol{\delta X}_+\!\!\int_{\mathbf{0}}^{\mathbf{0}}\!\!\mathcal{D}\boldsymbol{\delta X}_-
e^{\mi S_s[\delta \boldsymbol{X}_+,\boldsymbol{j}_+]-\mi S_s[\delta \boldsymbol{X}_-,\boldsymbol{j}_-]} \Big |_{\boldsymbol{j}_\pm = \boldsymbol{0}} \\
& = e^{\mi S_{_\mathrm{IF}}\left [\overline{\boldsymbol{X}}_{\alpha,+} - \mi \tfrac{\boldsymbol{\delta}}{\boldsymbol{\delta j}_+},\overline{\boldsymbol{X}}_{\beta,-}+ \mi \tfrac{\boldsymbol{\delta}}{\boldsymbol{\delta j}_-}\right ]}  \\
& \qquad \quad \qquad \quad \qquad  \times \left ( \tfrac{M}{2\pi(t_f-t_i)} \right )^3\!\underbrace{e^{\tfrac{\mi}{2}\!\!\int_{t_i}^{t_f}\!\!\mathrm{d}t\!\!\int_{t_i}^{t_f}\!\!\mathrm{d}t'\;
G_\mathrm{d}(t,t')\,\left[ \boldsymbol{j}_+(t)\cdot\boldsymbol{j}_+(t')-\boldsymbol{j}_-(t)\cdot\boldsymbol{j}_-(t')\right]}}_{Z_0[\boldsymbol{j_+},\boldsymbol{j_-}]} \Bigg |_{\boldsymbol{j}_\pm = \boldsymbol{0}} \, , \\
\end{aligned}
\end{equation}
where the generating functional for worldline fluctuations $Z_0[\boldsymbol{j}_+,\boldsymbol{j}_-]$ is automatically normalised, and we introduced the Dirichlet Green's function~\cite{Corradini:2015tik}
\begin{equation}\label{eq:Gd}
G_\mathrm{d}(t,t') = \frac{(t_{<}-t_{i})(t_{>}-t_f)}{M(t_f-t_i)} \,, \quad \, t_{<} = \min(t,t')\, , \quad   t_{>} = \max(t,t') \, .
\end{equation}

To make progress, note that the classical solution of Eq.~\eqref{eq:unperturbed} is a sum of the homogeneous (i.e., straight-line) and particular (i.e., sourced) solutions
\begin{equation}\label{eq:Xbar}
\overline{\boldsymbol{X}}_{\alpha,\pm}(t) \;=\; 
\mathbf{x}_{i,\pm}\,\frac{t_f-t}{t_f-t_i} \,+\, \mathbf{x}_f\,\frac{t-t_i}{t_f-t_i} \,+\, 
\underbrace{\int_{t_i}^{t_f}\!\mathrm{d}t'\;G_\mathrm{d}(t,t')\,\boldsymbol{J}_{\alpha,\pm}(t')}_{\overline{\boldsymbol{x}}_{\alpha,\pm}(t)} \,,
\end{equation}
where $\overline{\boldsymbol{x}}_{\alpha,\pm}(t)$ is the particular solution. 
In light of the closed and symmetric interferometric geometry (cf. Fig.~\ref{fig:paths}), i.e., $\boldsymbol{J}_{L}=-\boldsymbol{J}_{R}$, and for $t_f=-t_i=\mathcal{T} > T_s$, where $T_s = T+T'$ is the time-scale of the interferometric sequence, the classical action (cf. Eq.~\eqref{app:eq:Scl}) may be expressed as
\begin{equation}
S_s[\overline{\boldsymbol{X}}_{\alpha, \pm},\boldsymbol{J}_{\alpha, \pm}]\!=\!\!\tfrac{M|\mathbf{x}_f-\mathbf{x}_{i,\pm}|^{2}}{4\mathcal{T}}  + \tfrac{1}{2}\int_{-\mathcal{T}}^\mathcal{T}\!\mathrm{d}t\int_{-\mathcal{T}}^\mathcal{T}\!\mathrm{d}t' G_\mathrm{d}(t,t') \boldsymbol{J}_{\alpha, \pm}(t) \cdot \boldsymbol{J}_{\alpha, \pm}(t') \, .
\end{equation}
Consequently, the difference between the classical actions in the first line of Eq.~\eqref{app:eq:rhos} becomes independent of the particular solution. 
Using Eq.~\eqref{app:eq:step2} and recalling from Eq.~\eqref{eq:initial_state} that $\bra{\alpha,\mathbf{x}_{i,+}}\hat{\rho}_{_{s}}(t_i)\ket{\beta,\mathbf{x}_{i,-}} = \tfrac{1}{2}\Phi(\mathbf{x}_{i,+})\Phi^*(\mathbf{x}_{i,-})$,
\begin{equation}\label{app:eq:step}
\begin{aligned}
\text{Eq.~\eqref{app:eq:rhos}} & = \tfrac{1}{2} \left ( \tfrac{M}{4\pi\mathcal{T}} \right )^3 \int\!\!\mathrm{d}^3 x_{i,+}\int \mathrm{d}^3 x_{i,-} \!\!\, \Phi(\mathbf{x}_{i,+})\Phi^*(\mathbf{x}_{i,-})
\\ & \qquad \qquad \times
\int\!\!\mathrm{d}^3 x_f \, e^{\tfrac{\mi M}{4\mathcal{T}} \left (|\mathbf{x}_f-\mathbf{x}_{i,+}|^{2}-|\mathbf{x}_f-\mathbf{x}_{i,-}|^{2}\right)} e^{\mi S_{_\mathrm{IF}}\left [\overline{\boldsymbol{X}}_{\alpha,+} - \mi \tfrac{\boldsymbol{\delta}}{\boldsymbol{\delta j}_+},\overline{\boldsymbol{X}}_{\beta,-}+ \mi \tfrac{\boldsymbol{\delta}}{\boldsymbol{\delta j}_-}\right ]}  Z_0[\boldsymbol{j}_+,\boldsymbol{j}_-]\Bigg |_{\boldsymbol{j}_\pm = 0}\\ 
& = \tfrac{1}{2} \left ( \tfrac{M}{4\pi\mathcal{T}} \right )^3 \int\!\!\mathrm{d}^3 x_{i,+}\int \mathrm{d}^3 x_{i,-} \!\!\, \Phi(\mathbf{x}_{i,+})\Phi^*(\mathbf{x}_{i,-}) e^{\tfrac{\mi M}{4\mathcal{T}}(|\mathbf{x}_{i,+}|^2-|\mathbf{x}_{i,-}|^2)}\\
& \qquad \qquad
\times \int\!\!\mathrm{d}^3 x_f \, e^{\tfrac{\mi M}{2\mathcal{T}} \mathbf{x}_f\cdot \left (\mathbf{x}_{i,-} -\mathbf{x}_{i,+}\right)} \left %
(1+\mi S_{_\mathrm{IF}}\left [\overline{\boldsymbol{X}}_{\alpha,+} - \mi \tfrac{\boldsymbol{\delta}}{\boldsymbol{\delta j}_+},\overline{\boldsymbol{X}}_{\beta,-}+ \mi \tfrac{\boldsymbol{\delta}}{\boldsymbol{\delta j}_-}\right ] +\mathcal{O}(g^3) \right ) Z_0[\boldsymbol{j}_+,\boldsymbol{j}_-]\Bigg |_{\boldsymbol{j}_\pm = 0} \\
& \equiv \tfrac{1}{2} +  \tfrac{\mi}{2} \left \langle S_{_\mathrm{IF}}\left [\overline{\boldsymbol{X}}_{\alpha,+} - \mi \tfrac{\boldsymbol{\delta}}{\boldsymbol{\delta j}_+},\overline{\boldsymbol{X}}_{\beta,-}+ \mi \tfrac{\boldsymbol{\delta}}{\boldsymbol{\delta j}_-}\right ] \right \rangle Z_0[\boldsymbol{j}_+,\boldsymbol{j}_-]\Big |_{\boldsymbol{j}_\pm = 0} +\mathcal{O}(g^3) \, , 
\end{aligned}
\end{equation}
where $\langle \cdot \rangle$ is defined as the average over the boundary conditions of the worldlines. 

We will now proceed to finding explicit expressions for the averaged influence action. 
Given that the system currents explicitly depend on the system worldlines, and therefore on the worldline boundary conditions, it will be convenient to work in four-momentum space. 
We represent the three-momentum FT of the system currents (cf. Eqs.~\eqref{eq:rhocl}--\eqref{eq:rhoq}) as
\begin{equation}\label{eq:rho_FT}
\widetilde\varrho^{\mathrm{cl},\mathrm{q}}(\mathbf{q},t;\boldsymbol{X}_{\pm})=F_A (\mathbf{q})\,\widetilde{\mathcal{W}}^{\mathrm{cl},\mathrm{q}}(\mathbf{q},t;\boldsymbol{X}_{\pm})\,,
\end{equation}
where the worldline histories are captured by the functions
\begin{equation}\label{app:eq:W}
\begin{aligned}
\widetilde{\mathcal{W}}^\mathrm{cl}(\mathbf{q},t;\boldsymbol{X}_{\pm})\!&=\!\tfrac{1}{2} \left[
e^{-\mi\mathbf{q}\cdot\boldsymbol{X}_{\alpha,+}(t)}\!+\!e^{-\mi\mathbf{q}\cdot\boldsymbol{X}_{\beta,-} (t)}
\right] 
\, , \\
\widetilde{\mathcal{W}}^\mathrm{q}(\mathbf{q},t;\boldsymbol{X}_{\pm})\!&=\!
e^{-\mi\mathbf{q}\cdot\boldsymbol{X}_{\alpha,+}(t)}\!-\!e^{-\mi\mathbf{q}\cdot\boldsymbol{X}_{\beta,-} (t)}
\, .
\end{aligned}
\end{equation}
which define the vector
\begin{equation}\label{app:eq:vector}
\vec{\widetilde{\mathcal{W}}}(\mathbf{q},t;\boldsymbol{X}_{\pm})^\mathsf{T} =   \left (\widetilde{\mathcal{W}}^\mathrm{cl}(\mathbf{q},t;\boldsymbol{X}_{\pm}) \quad \widetilde{\mathcal{W}}^\mathrm{q}(\mathbf{q},t;\boldsymbol{X}_{\pm}) \right ) \, ,
\end{equation}
Hence, with the aid of Eq.~\eqref{eq:Sif}, 
\begin{equation}\label{app:eq:Sif_q}
\begin{aligned}
\text{2\textsuperscript{nd} term of Eq.~\eqref{app:eq:step}} &= 
-\tfrac{\mi}{4}\!\left(\tfrac{g}{\Lambda^D}\right)^{\!2}\!\! \int_{-\mathcal{T}}^{\mathcal{T}} \mathrm{d} t \!\!  \int_{-\mathcal{T}}^{\mathcal{T}} \mathrm{d} t' \!\!  \int \dbar^3 q 
|F_A(\mathbf{q})|^2 \\
& \times 
\left \langle \vec{\widetilde{W}}\left (-\mathbf{q},t;\overline{\boldsymbol{X}}_{\pm}\mp \mi \tfrac{\boldsymbol{\delta}}{\boldsymbol{\delta j}_\pm}\right)^{\mathsf T}
\,\widetilde{\mathsf{K}}(\mathbf{q},t-t')\,
\vec{\widetilde{W}}\left(\mathbf{q},t';\overline{\boldsymbol{X}}_{\pm}\mp \mi \tfrac{\boldsymbol{\delta}}{\boldsymbol{\delta j}_\pm} \right) \right \rangle Z_0[\boldsymbol{j}_+,\boldsymbol{j}_-]\Big |_{\boldsymbol{j}_\pm = 0} \, .
\end{aligned}
\end{equation}
In light of Eqs.~\eqref{app:eq:W}--\eqref{app:eq:vector}, we can factorize the basic worldline-dependent objects entering Eq.~\eqref{app:eq:Sif_q} into classical path dependent and worldline fluctuation dependent pieces: for $m,n\in\{+,-\}$, the relevant object is
\begin{equation}\label{app:eq:average1}
\left\langle
e^{-\mi\mathbf{q}\cdot\overline{\boldsymbol{X}}_{m}(t)}
e^{+\mi\mathbf{q}\cdot\overline{\boldsymbol{X}}_{n}(t')}
\right\rangle e^{\mi \theta_{mn}(t,t')} \, , \quad  e^{\mi \theta_{mn}(t,t')} \equiv e^{-m\mathbf{q}\cdot\tfrac{\boldsymbol{\delta}}{\boldsymbol{\delta j}_m(t)}}e^{n\mathbf{q}\cdot\tfrac{\boldsymbol{\delta}}{\boldsymbol{\delta j}_n(t')}} Z_0[\boldsymbol{j}_+,\boldsymbol{j}_-]\Big |_{\boldsymbol{j}_\pm = 0} \, .
\end{equation}
By definition,
\begin{equation}\label{app:eq:theta}
\begin{aligned}
\theta_{++}(t,t') = - \theta_{--}(t,t')
&=
\frac{\mathbf q^2}{2}
\left[
G_{\rm d}(t,t)
+
G_{\rm d}(t',t')
-
2G_{\rm d}(t,t')
\right],
\\
\theta_{+-}(t,t')
= -\theta_{-+}(t,t') &= 
\frac{\mathbf q^2}{2}
\left[
G_{\rm d}(t,t)
-
G_{\rm d}(t',t')
\right] \, ,
\end{aligned}
\end{equation}
and
\begin{equation}\label{app:eq:average}
\begin{aligned}
\left\langle
e^{-\mi\mathbf{q}\cdot\overline{\boldsymbol{X}}_{m}(t)}
e^{+\mi\mathbf{q}\cdot\overline{\boldsymbol{X}}_{n}(t')}
\right\rangle &= \left\langle
e^{-\mi\mathbf{q}\cdot\left[\mathbf{x}_{i,m}\left(\tfrac{\mathcal{T}-t}{2\mathcal{T}}\right) - \mathbf{x}_{i,n}\left(\tfrac{\mathcal{T}-t'}{2\mathcal{T}}\right) + \mathbf{x}_f\left ( \tfrac{t-t'}{2\mathcal{T}} \right )+\overline{\boldsymbol{x}}_{m}(t)-\overline{\boldsymbol{x}}_{n}(t')\right]}
\right\rangle \\
& =  e^{-\mi\mathbf{q}\cdot\left(\overline{\boldsymbol{x}}_{m}(t)-\overline{\boldsymbol{x}}_{n}(t')\right)} \int\!\!\mathrm{d}^3 x_{i,+}\int \mathrm{d}^3 x_{i,-} \!\!\, \Phi(\mathbf{x}_{i,+})\Phi^*(\mathbf{x}_{i,-}) \delta^{(3)}\!\left(
\mathbf{x}_{i,-}-\mathbf{x}_{i,+}
+\tfrac{t'-t}{M}\mathbf{q}
\right) \\
& \qquad \qquad \times e^{-\mi\mathbf{q}\cdot\left[\mathbf{x}_{i,m}\left(\tfrac{\mathcal{T}-t}{2\mathcal{T}}\right) - \mathbf{x}_{i,n}\left(\tfrac{\mathcal{T}-t'}{2\mathcal{T}}\right) \right]}e^{\tfrac{\mi M}{4\mathcal{T}}(|\mathbf{x}_{i,+}|^2-|\mathbf{x}_{i,-}|^2)} \\
& = e^{-\mi\mathbf{q}\cdot\left(\overline{\boldsymbol{x}}_{m}(t)-\overline{\boldsymbol{x}}_{n}(t')\right)} e^{\mi \mathcal{\phi}_{mn}(t,t')}\int\!\!\mathrm{d}^3 x_{i,+} \!\!\, \Phi(\mathbf{x}_{i,+})\Phi^*(\mathbf{x}_{i,+}+\tfrac{t-t'}{M}\mathbf{q}) \, ,
\end{aligned}
\end{equation} 
where, in going to the second line of Eq.~\eqref{app:eq:average}, we performed the integral over $\mathbf{x}_f$, which produces the delta function, and in the final step we introduced
\begin{equation}\label{app:eq:phi}
\phi_{mn}(t,t')
=
\frac{|\mathbf{q}|^2(t-t')}{2M\mathcal T}
\left[
\nu_n(\mathcal T-t')
-
\nu_m(\mathcal T-t)
\right]
-
\frac{|\mathbf{q}|^2(t-t')^2}{4M\mathcal T} \, ,
\qquad \text{with} \quad 
\nu_+=0 \, ,
\qquad
\nu_-=1 \, .
\end{equation}
Using Eqs.~\eqref{app:eq:theta}--\eqref{app:eq:phi}, Eq.~\eqref{app:eq:average1} can then be expressed compactly as
\begin{equation}\label{app:eq:averaged_final}
\left\langle
e^{-\mi\mathbf{q}\cdot\overline{\boldsymbol{X}}_{m}(t)}
e^{+\mi\mathbf{q}\cdot\overline{\boldsymbol{X}}_{n}(t')}
\right\rangle e^{\mi \theta_{mn}(t,t')} = e^{-\mi\mathbf{q}\cdot\left(\overline{\boldsymbol{x}}_{m}(t)-\overline{\boldsymbol{x}}_{n}(t')\right)} e^{\mi\varphi_{mn}(t,t')} \int\!\!\mathrm{d}^3 x_{i,+} \!\!\, \Phi(\mathbf{x}_{i,+})\Phi^*(\mathbf{x}_{i,+}+\tfrac{t-t'}{M}\mathbf{q})  \, ,
\end{equation}
with $\varphi_{mn}(t,t')=\phi_{mn}(t,t')+\theta_{mn}(t,t')$ and explicitly given by
\begin{equation}\label{app:eq:varphi}
\begin{aligned}
\varphi_{++}(t,t') &= -\varphi_{--}(t,t') = -\tfrac{|\mathbf{q}|^2}{2M}|t-t'| \, , \\
\varphi_{+-}(t,t') &= -\varphi_{-+}(t,t') = \tfrac{|\mathbf{q}|^2}{2M}(t-t') \, .
\end{aligned}
\end{equation}

The heavy-probe limit should now be taken at fixed $\mathcal{T}$, $t$, and $t'$. 
In this limit, $\theta_{mn} \rightarrow 0$, the two initial worldline configurations are identified and, for normalized wavepackets $\Phi(\mathbf{x})$, the heavy probe limit of Eq.~\eqref{app:eq:averaged_final} is 
\begin{equation}
\lim_{M\rightarrow \infty}\left\langle
e^{-\mi\mathbf{q}\cdot\overline{\boldsymbol{X}}_{m}(t)}
e^{+\mi\mathbf{q}\cdot\overline{\boldsymbol{X}}_{n}(t')}
\right\rangle e^{\mi \theta_{mn}(t,t')} = e^{-\mi \mathbf{q}\cdot(\overline{\boldsymbol{x}}_m(t)-\overline{\boldsymbol{x}}_n(t'))} \, ,
\end{equation}
such that the driven classical histories remain encoded in the influence action via the particular solution $\overline{\boldsymbol{x}}_\pm$ (cf. Eq.~\eqref{eq:Xbar}).
Taking the asymptotic limit $\mathcal{T}\rightarrow \infty$, Eq.~\eqref{app:eq:Sif_q} simplifies to
\begin{equation}\label{app:eq:Sif_lim}
\lim_{\mathcal{T}\rightarrow \infty}\lim_{M\rightarrow \infty}
\text{Eq.~\eqref{app:eq:Sif_q}}= 
-\tfrac{\mi}{4}\!\left(\tfrac{g}{\Lambda^D}\right)^{\!2}\!\! \int \dbar^4 q \, |F_A(\mathbf{q})|^2 \,
\vec{\widetilde{W}}(-q;\overline{\boldsymbol{x}}_\pm)^{\mathsf T}
\,\widetilde{\mathsf{K}}(q)\,
\vec{\widetilde{W}}(q;\overline{\boldsymbol{x}}_\pm) + \mathcal{O}(g^3) \equiv \tfrac{\mi}{2} S_\mathrm{IF}[\overline{\boldsymbol{x}}_{\alpha},\overline{\boldsymbol{x}}_{\beta}] \, ,
\end{equation}
where $\widetilde{W}^\mathrm{q,cl}(q)$ are functions of four-momentum $q$ and $\overline{\boldsymbol{x}}_\pm$.
Hence, the system's reduced density matrix in the heavy-probe and asymptotic time limit takes the form  
\begin{equation}
\lim_{\mathcal{T}\rightarrow\infty}\lim_{M\rightarrow \infty}\rho_s^{\alpha \beta}(\mathcal{T}) = \tfrac{1}{2}  e^{\mi S_\mathrm{IF}[\overline{\boldsymbol{x}}_{\alpha},\overline{\boldsymbol{x}}_{\beta}]} + \mathcal{O}(g^3) \, ,
\end{equation}
with $S_\mathrm{IF}[\overline{\boldsymbol{x}}_{\alpha},\overline{\boldsymbol{x}}_{\beta}]$ given by Eq.~\eqref{app:eq:Sif_lim}. 

\subsection{Beyond the heavy-probe limit}

To go beyond the heavy-probe limit one needs to keep $M$ finite. 
Physically, this corresponds to accounting for recoil or response corrections of the atom's worldline as a result of the system's coupling to the environment. 
As a result, deviations from the classical unperturbed trajectory can be organized as a series expansion in $1/M$. 

As is clear from Eqs.~\eqref{app:eq:averaged_final}--\eqref{app:eq:varphi}, finite mass corrections enter the observable via $\varphi_{mn}(t,t')$ and $1/M$ terms in the arguments of the initial wavepackets. In the former, the correction is a phase $\tfrac{|\mathbf{q}|^2}{2M}(t-t')$. Since the system currents have support for $t,t' \in [-T_s/2,T_s/2]$, where $T_s$ is the interferometer time, and $\widetilde{\mathsf{K}}(\mathbf{q};t-t')$ has support for $|t-t'| \lesssim \tau_e$, where $\tau_e$ is the time-scale of the environment (e.g., the coherence time for quadratically coupled DM), these corrections are parametrically $|\mathbf{q}|^2 \tau_\mathrm{eff}/M$, where  $\tau_\mathrm{eff} = \min(T_s,\tau_e)$. 
We interpret this as the eikonal phase acquired by an atom whose center-of-mass trajectory recoils under a momentum transfer $\mathbf q$, with recoil velocity $\mathbf v_\mathrm{rec}=\mathbf q/M$. For Gaussian wavepackets with width $\sigma$, the latter correction can be seen from
\begin{equation}
\int \mathrm{d}^3 x_{i,+} \Phi(\mathbf{x}_{i,+}) \Phi^*(\mathbf{x}_{i,+}+\mathbf{q}\tfrac{t-t'}{M}) = e^{-|\mathbf{q}|^2(t-t')^2/8M^2\sigma^2} \, .
\end{equation}
This correction scales as $|\mathbf v_{\rm rec}\tau_{\rm eff}|^2/\sigma^2$ and measures the sensitivity of the initial c.o.m. state to the recoil-induced displacement.  It is conceptually distinct from the form factor $F_A(\mathbf q)$: the former is a property of the prepared wavepacket, while the latter is a property of the atom's internal charge distribution.

\section{Derivation of the decoherence and phase shift observables}\label{sec:geom}

For the interferometric sequence depicted in Fig.~\ref{fig:paths}, we take $\boldsymbol{J}_R\!=\!-\boldsymbol{J}_L$ with
\begin{align}
\boldsymbol{J}_L(t)\!=\!\tfrac{M\mathbf{v}}{2}\!&\Big[\!-\!\delta\left(t\!+\!\tfrac{T+T'}{2}\right)\!+\!\delta\left(t\!+\!\tfrac{T}{2}\right)
\!+\!\delta\left (t\!-\!\tfrac{T}{2}\right)\!-\!\delta \left (t\!-\!\tfrac{T+T'}{2}\right)\Big] \, ,
\end{align}
so that $|\boldsymbol{\Delta x}|=|\mathbf{v}|T'/2$, where $|\mathbf{v}|$ is the maximum relative speed between the two arms. 

As we showed earlier the influence action in the heavy probe and asymptotic time limit depends exclusively on the particular solution (cf. Eq.~\eqref{app:eq:Sif_lim}), which takes the asymptotic time form 
\begin{equation}
\lim_{\mathcal{T}\rightarrow \infty} \lim_{M\rightarrow \infty} \overline{\boldsymbol{x}}_{\alpha}(t) =
\int\!\mathrm{d}t'\;\tfrac{|t-t'|}{2M}\,\boldsymbol{J}_{\alpha}(t') \, .
\end{equation}
Hence, using Eq.~\eqref{eq:rho_FT}, the appropriate FT of the system currents in the regime $T\gg T'$ evaluates to
\begin{equation}\label{app:eq:syscurrent}
\!\!\!\!\int\!\mathrm{d}^4x\,\varrho(\mathbf{x}\!-\!\overline{\boldsymbol{x}}_{{L,R}}(t))e^{\mi q\cdot x}\!\approx \!F_A(\mathbf{q})\bigl[2\pi\delta(q_0)+T\,\mathrm{sinc}(q_0T/2)(e^{\pm\mi\mathbf{q}\cdot\boldsymbol{\Delta x}/2}-1)\bigr] \, .
\end{equation}

Equipped with Eqs.~\eqref{eq:rho_FT}, \eqref{app:eq:W}, \eqref{app:eq:Sif_lim} and ~\eqref{app:eq:syscurrent},  we can compute the Keldysh-basis bilinears that appear in Eq.~\eqref{eq:rho_heavy}: 
\begin{align}
\widetilde{W}_\mathrm{cl}(q;\overline{\boldsymbol{x}}_\pm)\widetilde{W}_\mathrm{q}^*(q;\overline{\boldsymbol{x}}_\pm) &\!=\!-4\mi T\sin\!\Bigl(\tfrac{\mathbf{q}\cdot \boldsymbol{\Delta x}}{2}\Bigr)
\!\Bigl [\!\pi\delta(q_0)\!-\!T\,\mathrm{sinc}^2\!\Bigl(\tfrac{q_0T}{2}\Bigr)\sin^2\!\Bigl(\tfrac{\mathbf{q}\cdot\boldsymbol{\Delta x}}{4}\Bigr)\!\Bigr ] \, , \\[2pt]
|\widetilde{W}_\mathrm{q}(q;\overline{\boldsymbol{x}}_\pm)|^2 &=\!2T^2\mathrm{sinc}^2\!\Bigl(\tfrac{q_0T}{2}\Bigr)\bigl[1\!-\!\cos(\mathbf{q}\!\cdot\!\boldsymbol{\Delta x})\bigr] \, .
\end{align}
Multiplying by the appropriate environment kernels yields the phase shift and decoherence observables (cf. Eqs.~\eqref{eq:P_final} and \eqref{eq:D_final}, respectively).

\section{Environment kernels}\label{sec:kernels}

For the operators $O_\chi=m_\chi\chi^2$ (real scalar) and $O_\chi=\overline\chi\chi$ (Dirac fermion), the SK kernels are obtained from standard one-loop diagrams~\cite{Badurina:2026myo,Kamenev_2011}. Using the definitions in Eqs.~\eqref{eq:ret}-\eqref{eq:kel} and assuming DM in a Gaussian state, the retarded and Keldysh kernels in four momentum space can be expressed in terms of two-point functions of $\chi$ as
\begin{equation}
\begin{aligned}
\widetilde{K}_{\mathrm{r}}(q) =
\begin{cases}
\displaystyle 4 \mi m_\chi^2 \!\! \int \dbar^4 p \,   \widetilde{G}^{^\chi}_{\mathrm{k}}(p)\widetilde{G}^{^\chi}_{\mathrm{r}}(q-p) & \text{for a real scalar} \, ,\\
\displaystyle -\frac{\mi}{2} \int \dbar^4 p \,\mathrm{tr} \big [\widetilde{G}^{^\chi}_{\mathrm{r}}(p+q) \widetilde{G}^{^\chi}_{\mathrm{k}}(p) + \widetilde{G}^{^\chi}_{\mathrm{a}}(p)\widetilde{G}^{^\chi}_{\mathrm{k}}(p+q) \big ] &  \text{for a Dirac fermion} \, ,
\end{cases}
\end{aligned}
\end{equation}
and
\begin{equation}
\widetilde{K}_{\mathrm{k}}(q) =
\begin{cases}
 \displaystyle 2 \mi m_\chi^2 \!\!\int \dbar^4 p  \left [\widetilde{G}^{^\chi}_{\mathrm{k}}(p)\widetilde{G}^{^\chi}_{\mathrm{k}}(q-p) +\tfrac{1}{4}\widetilde{G}^{^\chi}_{\mathrm{r}}(p) \widetilde{G}^{^\chi}_{\mathrm{r}}(q-p)+\tfrac{1}{4}\widetilde{G}^{^\chi}_{\mathrm{a}}(p) \widetilde{G}^{^\chi}_{\mathrm{a}}(q-p) \right ]  & \text{for a real scalar} \, , \\
\displaystyle -\frac{\mi}{4} \!\int \dbar^4 p \,\mathrm{tr} \big [ \widetilde{G}^{^\chi}_{\mathrm{k}}(p) \widetilde{G}^{^\chi}_{\mathrm{k}}(p+q) +\widetilde{G}^{^\chi}_{\mathrm{r}}(p) \widetilde{G}^{^\chi}_{\mathrm{a}}(p+q)+\widetilde{G}^{^\chi}_{\mathrm{a}}(p) \widetilde{G}^{^\chi}_{\mathrm{r}}(p+q) \big ] & \text{for a Dirac fermion} \, ,
\end{cases}
\end{equation}
where the trace is over spinor indices. The retarded Green's functions of $\chi$ in four-momentum representation are
\begin{equation}
\begin{aligned}
\widetilde{G}^{^\chi}_{\mathrm{r}}(p) = \begin{cases}
\dfrac{1}{(p_0 + \mi \epsilon)^2-E_{\mathbf{p}}^2} = \mathrm{P}\left [\dfrac{1}{p_0^2-E_{\mathbf{p}}^2} \right] + \dfrac{\mi}{4E_\mathbf{p}}\left[\deltabar(p_0+E_\mathbf{p})-\deltabar(p_0-E_\mathbf{p})\right] & \text{for a real scalar} \, ,\\ \dfrac{\slashed{p}+m_\chi}{(p_0 + \mi\epsilon)^2-E_\mathbf{p}^2} = (\slashed{p}+m_\chi)\left \{ \mathrm{P}\left [\dfrac{1}{p_0^2-E_{\mathbf{p}}^2} \right] + \dfrac{\mi}{4E_\mathbf{p}}\left[\deltabar(p_0+E_\mathbf{p})-\deltabar(p_0-E_\mathbf{p})\right]\right \} & \text{for a Dirac fermion} \, ,
\end{cases}
\end{aligned}
\end{equation}
where $\mathrm{P}[\cdot]$ is the principal value, $\widetilde{G}^{^\chi}_{\mathrm{a}}(p)$ is given by $\widetilde{G}^{^\chi}_{\mathrm{r}}(p)$ under the map $\mi \epsilon\to-\mi \epsilon$ and we used the Larkin-Ovchinnikov convention for fermions~\cite{Larkin_Ovchinnikov_1975}. With our set of conventions, the Keldysh propagators can be expressed as
\begin{equation}
\begin{aligned}
\widetilde{G}^{^\chi}_{\mathrm{k}}(p) = - \frac{\mi}{4E_\mathbf{p}} 
\begin{cases}
\displaystyle \big[(1+2n_{_\chi}(\mathbf{p}))\deltabar(p_0-E_\mathbf{p}) + (1+2n_{_\chi}(-\mathbf{p}))\deltabar(p_0+E_\mathbf{p}) \big] \!\!\ & \!\!\  \text{for a real scalar} \, , \\ 
\displaystyle  2 (\slashed{p}+m_{_\chi}) \big [ \deltabar(p_0-E_\mathbf{p}) (1-n_{_\chi}(\mathbf{p})) + \deltabar(p_0+E_\mathbf{p}) \big ]\!\!\ & \!\!\ \text{for a Dirac fermion} \, ,
\end{cases}
\end{aligned}
\end{equation}
where $n_\chi(\mathbf{p}) = \tfrac{\rho_{_\mathrm{DM}}}{m_\chi} f_\chi(\mathbf{p})$ is the (spin-summed) occupation number per three-momentum $\mathbf{p}$, given the local DM energy density $\rho_{_\mathrm{DM}}$ and the DM three-momentum distribution $f_\chi(\mathbf{p})$. Note that, for fermionic DM, we have assumed an unpolarized DM background. In this case, $n_\chi(\mathbf{p})\leq 2$ and we set the antiparticle occupation number to zero.  
The Bose enhancement and Pauli blocking factors appear \emph{only} in the noise kernel because the retarded kernel is built from a commutator. This is the SK origin of the sharp dichotomy between $\mathbb{D}$ and $\mathbb{P}$ identified in the main text.

For linearly-coupled DM, it immediately follows that $\widetilde{K}_{\mathrm{r}}(q) = \widetilde{G}^{^\chi}_{\mathrm{r}}(q)$ and $\widetilde{K}_{\mathrm{k}}(q) = \widetilde{G}^{^\chi}_{\mathrm{k}}(q)$; hence only $\widetilde{K}_{\mathrm{k}}(q)$ receives $n_\chi$-dependent effects.
\end{document}